\documentclass[reprint,aps,twocolumn, superscriptaddress]{revtex4-1}

\usepackage[hidelinks]{hyperref} 
\usepackage{graphicx}
\usepackage{amsmath}
\usepackage{amssymb}
\usepackage{siunitx}
\usepackage{braket}

\DeclareGraphicsExtensions{.png,.jpg,.eps}
\usepackage{xcolor}
\hypersetup{
    colorlinks,
    linkcolor={red!50!black},
    citecolor={blue!80!black},
    urlcolor={blue!80!black}
}

\def\m1{{^{-1}}}

\begin{document}

\author{Bastian Zinkl}
\affiliation{Institute for Theoretical Physics, ETH Zurich, 8093 Zurich, Switzerland}

\author{Mark H. Fischer}
\affiliation{Institute for Theoretical Physics, ETH Zurich, 8093 Zurich, Switzerland}
\affiliation{Department of Physics, University of Zurich, 8057 Zurich, Switzerland}

\author{Manfred Sigrist}
\affiliation{Institute for Theoretical Physics, ETH Zurich, 8093 Zurich, Switzerland}

\title{
Superconducting gap anisotropy and topological singularities due to lattice translational symmetry and their thermodynamic signatures 
}

\begin{abstract}
Symmetry arguments based on the point group of a system and thermodynamic measurements are often combined to identify the order parameter in unconventional superconductors.
However, lattice translations, which can induce additional momenta with vanishing order parameter in the Brillouin zone, are neglected, especially in gap functions otherwise 
expected to be constant, such as in chiral superconductors.
After a general analysis of the symmetry conditions for vanishing gap functions, we study the case of chiral $p$- and chiral $f$-wave pairing on a square lattice, a situation relevant for Sr$_2$RuO$_4$.
Specifically, we calculate the impurity-induced density of states,
specific heat, superfluid density and thermal conductivity employing a
self-consistent T-matrix calculation and compare our results to the case of a nodal ($d$-wave) order parameter.
While there is a clear distinction between a fully
gapped chiral state and a nodal state, the strongly anisotropic case is
almost indistinguishable from the nodal case.
Our findings illustrate the difficulty of
interpreting thermodynamic measurements. In particular, we find
that the available measurements are consistent with a chiral ($f$-wave) order
parameter.
Our results help to reconcile the thermodynamic measurements 
with the overall picture of chiral spin-triplet superconductivity in Sr$_2$RuO$_4$.

\end{abstract}

\date{\today}

\maketitle


\section{Introduction}
Generalizing unconventional superconductivity from the isotropic case to a lattice is usually done through the analysis of the generating point group. While the resulting classification is in terms of the irreducible representations of the point group, the nomenclature of the isotropic case, in other words in terms of angular momentum, is still generally used for simplicity and corresponds to the long-wavelength behavior of the order parameter. A result of such a symmetry analysis is the imposition of zeros in the gap function, resulting in a nodal structure of the excitation spectrum, with point or line nodes, depending on the symmetry properties under rotations or mirror operations. In two dimensions, for example, a $d$-wave order parameter has point nodes on either both axes or diagonals, while a $p$-wave order parameter only has point nodes on one axis. However, combining order parameters of a two-dimensional irreducible representation into a chiral state, these nodes can be lifted and a fully gapped state would be expected.

Experimentally, the nodal structure of a superconductor and the resulting low-energy excitations lead to distinct low-temperature thermodynamic responses, in particular power-law behavior in the temperature. These include the specific heat, superfluid density or London penetration depth, and thermal conductivity, all of which are usually employed to characterize the superconducting state and its symmetry. 

However, the above described analysis neglects another set of symmetries of a lattice, namely translation symmetry. Translations in real space lead to translational symmetry also in momentum space and thus, the Brillouin zone (BZ). Furthermore, the translation symmetries in momentum space imprint further symmetry requirements onto the order parameter, which can result in additional zeros in the gap function as defined over the whole BZ. This is particularly important in systems, where no other nodes are expected, such as chiral superconductors.

Such considerations have come into focus in Sr$_2$RuO$_4$, 
where strong evidence for unconventional superconductivity has been 
gathered~\cite{maeno1994superconductivity, maeno2001maeno}. Early on, the superconducting state was predicted to be
of chiral $p$-wave symmetry \cite{rice1995sr2ruo4}, characterized by the $d$ vector 
$\mathbf{d}(\mathbf{k}) = \boldsymbol{\hat{z}} \Delta(T) (\sin k_x \pm i \sin k_y)$ 
\cite{mackenzie2003superconductivity, maeno2011evaluation, kallin2012chiral, ishida1998spin, luke1998time, 
luke2000unconventional, duffy2000polarized, nelson2004odd, xia2006high, ishida2015spin}.
Still, several discrepancies exist \footnote{
Note that due to recent Knight shift measurements \cite{pustogow2019pronounced}, the assumed spin-triplet pairing symmetry 
of the superconducting state of Sr$_2$RuO$_4$ has been questioned once again.}, amongst them low-temperature measurements of 
thermodynamic quantities, such as specific heat \cite{nishizaki1999effect, nishizaki2000changes, deguchi2004gap} 
and the London penetration depth \cite{bonalde2000temperature}, or thermal conductivity measurements 
\cite{tanatar2001thermal, tanatar2001anisotropy, izawa2001superconducting, hassinger2017vertical}. These measurements all indicate nodes in the superconducting order parameter and 
thus seemingly contradict the proposal of an intrinsically nodeless $p$-wave state in Sr$_2$RuO$_4$.
Note, however, that several theoretical works predict a substantial contribution of higher-angular-momentum spin-triplet states in Sr$_2$RuO$_4$, which manifest in strong gap anisotropies~\cite{raghu2010hidden, wang2013theory, scaffidi2014pairing}.

Motivated by the general observation of gap anisotropies unexpected by pure point group analysis and the specific case of Sr$_2$RuO$_4$, we analyze here 
the occurrence of gap anisotropy in chiral superconductors and its connection to the thermodynamic response 
in the presence of disorder. For this purpose, we employ a tight-binding description of a two-dimensional lattice and calculate microscopically the density of states, 
specific heat, superfluid density and thermal conductivity.
The disorder consists of non-magnetic impurities and the respective scattering is assumed to be isotropic and unitary, 
which allows us to employ a self-consistent T-matrix approximation.

To analyze all translation-symmetry-imposed zeros, we include nearest-, as well as next-nearest-neighbor interactions. This leads to an almost isotropic ($p$-wave)  
and a strongly anisotropic ($f$-wave) pairing gap. For comparison, we further add a nodal spin-singlet order parameter ($d$-wave). 
We observe that the strongly anisotropic case is hardly distinguishable from the nodal case once disorder is added.

The present article is structured as follows. In Sec.~\ref{sec: origin} we first discuss the origin and properties of symmetry-imposed zeros in a chiral gap function. We elaborate on the resulting gap anisotropy for the lowest order spin-triplet states on the square lattice. We further analyze the movability of the zeros within the framework of Ginzburg-Landau (GL) theory. In Sec.~\ref{sec: disorder} we add disorder to our system and examine how the superconducting state is affected. Furthermore, in Sec.~\ref{sec: td}, we study the influence of the gap anisotropy on several thermodynamic properties. 
Finally, we discuss our findings and conclude in Sec.~\ref{sec: conclusion}.

\section{Origin and properties of nodes in the chiral spin-triplet state} \label{sec: origin}

\subsection{General Symmetry considerations}
Throughout this paper, we use the following nomenclature: Defining the gap function over the whole BZ, we refer to \emph{zeros}, where the gap function vanishes. The transformation properties of a gap function under the symmetry operations of the generating point group can enforce zeros along high-symmetry lines or at high-symmetry points. \emph{Nodes}, on the other hand, denote zeros in the excitation spectrum, in other words zeros of the gap function coinciding with the normal state Fermi surface.

Given the point group $\mathcal{G}$ of the underlying lattice, the superconducting order parameter transforms under an element $g\in\mathcal{G}$ as
\begin{equation}
	g:\Delta_{\bf k} \mapsto \hat{U}_g \Delta_{U_g^{-1}\bf k} \overset{!}= e^{i\phi_g}\Delta_{\mathbf{k}},
	\label{eq:trafos}
\end{equation}
where $U_g$ and $\hat{U}_g$ are representations of $g$ in \textit{k} space and the space of the gap matrix, usually spin or orbital space. The right-hand side of Eq.~\eqref{eq:trafos} with $\phi_g\in [0, 2\pi)$ follows from the fact that $\Delta_{\mathbf{k}}$ transforms as an irreducible representation of $\mathcal{G}$~\footnote{
Note that in case of a higher-dimensional irreducible representation, the actual gap function breaks additional symmetries and thus always transforms as a one-dimensional irreducible representation of the remaining symmetry group.}. For any non-trivial phase $\phi_g$, this transformation behavior defines zeros in the 
gap function for any ${U_g^{-1}\bf k}={\bf k}$. Examples include the vanishing of the gap function at the $\Gamma$ point for any angular-momentum channel $l\neq 0$ with $g$ a rotation or along the 
main axis or diagonals for a $d$-wave order parameter with $g$ a mirror operation.

A lattice, however, obbeys additional symmetries, namely translations. In addition to the ones discussed above, Eq.~\eqref{eq:trafos} then indicates further zeros in the gap function for any momentum satisfying 
\begin{equation}
	U_g^{-1}\mathbf{k} + \mathbf{G} = \mathbf{k}
	\label{eq:morezeros}
\end{equation}
with $\mathbf{G}$ a reciprocal lattice vector if $\phi_g \neq 0$. To name an example, this condition implies a vanishing order parameter on the BZ boundary for the case of $d_{xy}$-wave pairing on a square lattice. More interestingly, the condition implies points of zeros for chiral superconductors as we will discuss in the following using specific lattice realizations.

To illustrate the appearance of zeros in the gap function and the most important features in the topological nature of the chiral superconducting phase, we first introduce a tight-binding model on the square lattice.
The pairing interaction is formulated on the lattice, too, with nearest- and next-nearest-neighbor pairing. While we use this model primarily to study the chiral spin-triplet order parameters, 
it can also be used to study a nodal spin-singlet gap function for comparison in Sec.~\ref{sec: td}.
Due to the lattice formulation, the structure of the gap function is adapted to the Brillouin zone (BZ) with a corresponding nodal structure. Note that the same arguments apply to lattices with a three-fold rotation axis, as we will discuss at the end of this section.

\subsection{Two-dimensional square-lattice model}

In the following, we study a system described by a Hamiltonian including nearest-neighbor (NN) and next-nearest-neighbor
(NNN) hopping,
\begin{equation}
	\mathcal{H} =  \sum_{i, j, s} t_{i j} a^{\dagger}_{i, s} a^{\phantom{\dagger}}_{j, s} + V_{\text{pair}}, \label{eqn: eq2}
\end{equation}
where $a^{\dagger}_{i, s}$ ($a^{\phantom{\dagger}}_{i, s}$) denotes the 
creation (annihilation) operator for an electron with 
spin $s = \uparrow, \downarrow$ on site $i = (x_i, y_i)$. 
The dispersion in $k$ space is then given by 
\begin{align}
	\xi_{\mathbf{k}} = -2 t_1 ( \cos k_x + \cos k_y ) - 4 t_2 \cos k_x \cos k_y - \mu, 
\end{align}
where we have introduced the chemical potential $\mu$ and the first BZ is given by $k_x, k_y \in [-\pi, \pi]$ taking the lattice constant $a=1$.
In the following, we set the NN-hopping $t_1=1$ and the NNN-hopping $t_2 = 0.3$.
We restrict the pairing potential $V_{\text{pair}}$ 
to the spin-triplet channel with (in-plane) equal-spin pairing,
\begin{align}
	V_{\text{pair}} = \sum_{\substack{i, j \\ s_1 \dots s_4 }} V_{i, j}\
		\sigma_{s_1 s_2}^{x} \sigma_{s_3 s_4}^{x} a^{\dagger}_{i, s_1} a^{\dagger}_{j, s_2} a^{\phantom{\dagger}}_{j, s_3} a^{\phantom{\dagger}}_{i, s_4},  
\end{align}
where $\sigma^x$ is a Pauli matrix, $V_{i j}=V_1$ for NN-pairing, and $V_{ij}=V_2$ for NNN-pairing.
In momentum space, we obtain 
\begin{align}
	V_{\text{pair}} = \sum_{\substack{\mathbf{k}, \mathbf{k'} \\ s_1 s_2 }} 
		V_{\mathbf{k}\mathbf{k'}}\  
		c^{\dagger}_{\mathbf{k}, s_1} c^{\dagger}_{-\mathbf{k}, -s_1} 
		c^{\phantom{\dagger}}_{-\mathbf{k'}, -s_2} c^{\phantom{\dagger}}_{\mathbf{k'}, s_2},  
\end{align}
where 
\begin{align}
	V_{\mathbf{k}\mathbf{k'}} = \sum_{a = x, y}  \left[ V_1 \Phi_{1a}(\mathbf{k})\Phi_{1a}(\mathbf{k'}) 
	+ V_2 \Phi_{2a}(\mathbf{k})\Phi_{2a}(\mathbf{k'}) \right], \label{eqn: pairpot}
\end{align}
with basis funtions
\begin{align}
	\Phi_{1x}(\mathbf{k}) &= \sin k_x, \quad \Phi_{1y}(\mathbf{k}) = \sin k_y, \label{phi1}\\[1mm]
	\Phi_{2x}(\mathbf{k}) &= \sin k_x \cos k_y, \quad \Phi_{2y}(\mathbf{k}) = \sin k_y \cos k_x. \label{phi2}
\end{align}
These represent the simplest odd-parity pairing states 
on a square lattice with $\Phi_{ia}(-\mathbf{k}) = -\Phi_{ia}(\mathbf{k})$, $i=1,2$.

Within standard mean-field theory, the pairing interaction results in the quasiparticle gap
function 
\begin{multline}
	\Delta_{\mathbf{k}} = \Delta_p (\sin k_x \pm i \sin k_y) \\
	 +\Delta_f (\sin k_x \cos k_y \pm i \sin k_y \cos k_x ), \label{gap-1}
\end{multline}
which is a time-reversal-symmetry breaking phase with
chiral character. For convenience, we refer to the NN part as $p$-wave
and the NNN part as $f$-wave component. The coefficients 
$\Delta_{p, f}$ are obtained through the solution of 
the self-consistency equation \cite{mineev1999introduction}, 
\begin{align}
	\Delta_{\mathbf{k}} = - T \sum_{n} \sum_{\mathbf{k'}} V_{\mathbf{k}\mathbf{k'}} 
	\frac{\Delta_{\mathbf{k'}}}{\omega_n^2 + \xi_{\mathbf{k'}}^2 + \Delta_{\mathbf{k'}}^2}, \label{eqn: sc-gap}
\end{align}
where $V_{\mathbf{k}\mathbf{k'}}$ is the pairing potential in momentum space for the spin-triplet channel, 
$\omega_n = (2 n + 1)\pi k_B T$ are the fermionic Matsubara frequencies and $T$ denotes the temperature.
Using the full gap $\Delta_{\mathbf k}$, Eq.~\eqref{gap-1}, we can trace out the two basis 
functions $\Phi_{ix}({\mathbf k})$ to find two coupled equations, one for $\Delta_p$ and one for $\Delta_f$,
\begin{align}
	\begin{pmatrix}
\Delta_p \\
\Delta_f 
\end{pmatrix} = \sum_{\mathbf{k}} \mathcal{C}_{\mathbf{k}}
\begin{pmatrix}
V_1 & V_2 \cos k_y \\
V_1 \cos k_y & V_2 \cos^2 k_y  
\end{pmatrix}
\begin{pmatrix}
\Delta_p \\
\Delta_f 
\end{pmatrix} .
\end{align}
The factor $\mathcal{C}_{\mathbf{k}}$ is given by
\begin{align}
	\mathcal{C}_{\mathbf{k}} = -T \sum_{n} \frac{\sin^2 k_x}{\tilde{\omega}_n^2 + \xi_{\mathbf{k}}^2 + |\Delta_{\mathbf k}|^2} .
\end{align}
For a given strength of the pairing potentials for $p$- and $f$-wave pairs, $V_1$ and $V_2$,  
we can self-consistently tune the values of $\Delta_p$ and $\Delta_f$.



\subsection{Singularities and topological properties of the gap function}

The NN- and NNN-pairing states in Eq.~\eqref{phi1} and \eqref{phi2} have to satisfy several symmetry properties due to underlying lattice symmetries. 
In App.~\ref{sec:singularities}, we use the previous example of spin-triplet pairing on a square lattice to 
show in detail how the singularities of the gap function follow from general symmetry considerations alone. In this section however, we start the 
discussion directly with the quasiparticle gap function [Eq.~\eqref{gap-1}].

\begin{figure}[t!]
 \centering
	\includegraphics[width=0.65\columnwidth]{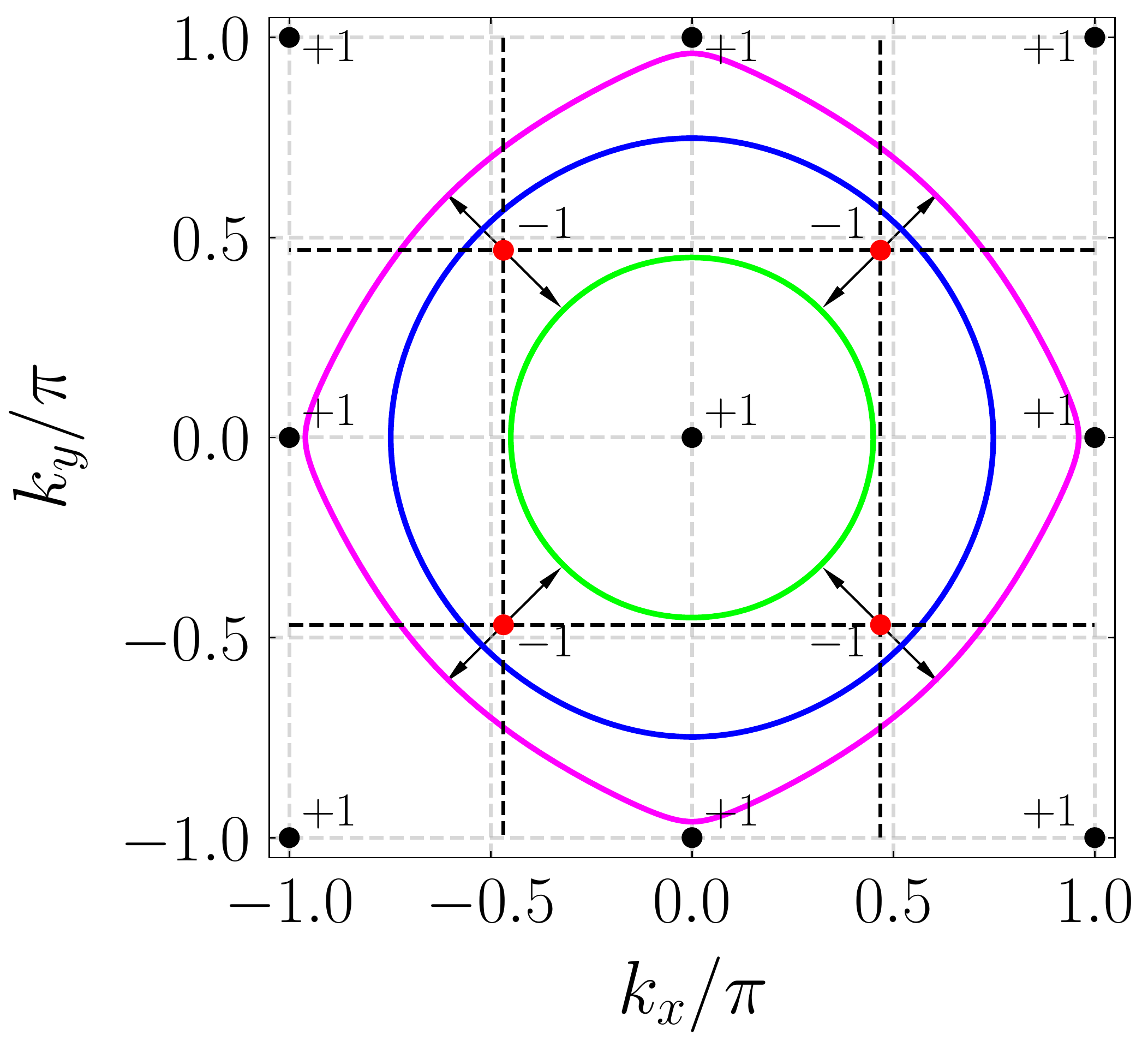}
	\caption{
		Example Fermi surfaces for $\mu=-1.0$ (green), 
		$\mu=0.1$ (blue) and $\mu=0.47$ (pink).
		The black dots mark the symmetry imposed gap zeros with associated winding numbers for dominant $f$-wave order parameter and negative chirality. 
		For a superposition of $p$-wave and $f$-wave pairs with $|\Delta_p| < |\Delta_f|$, the zeros move along the diagonals towards the BZ corner (center),
		for equal (different) relative sign.\\
	}
	\label{fig:nodes}
\end{figure}

To each of the zeros in $\Delta_{\mathbf{k}}$ we can attribute a topological charge corresponding to the winding number
\begin{equation}
N_{C} = \frac{1}{2 \pi}  \oint_C d\mathbf{k} \cdot \mathbf{\nabla}_{\mathbf{k}} \theta_{\mathbf k}  \label{chern},
\end{equation}
where $ \theta_{\mathbf k} = \arg [\Delta_{\mathbf{k}}] $ and $C$ denotes a closed path in positive direction around the zero without enclosing any other zero.
Considering $p$- and $f$-wave order parameters, all zeros possess a charge of either $+1$ or $-1$. The relations (\ref{Q-inter}) and (\ref{Q-intra}) connect zeros of the same charge. 
The zeros and their charge are shown in Fig.~\ref{fig:nodes}. Note that the charges change sign under time reversal operation.

The charges defined above are connected to the topology of the superconducting phase. For chiral superconducting states in two dimensions, the topological invariant is the Chern number, which is defined through Eq.~\eqref{chern} by taking the (normal-state) Fermi surface as the path $C$. The Chern number is then the sum over all charges encircled by the Fermi surface. 

\begin{figure}[t!]
 \centering
	\begin{minipage}[t]{0.49\columnwidth}
			\includegraphics[width=\textwidth]{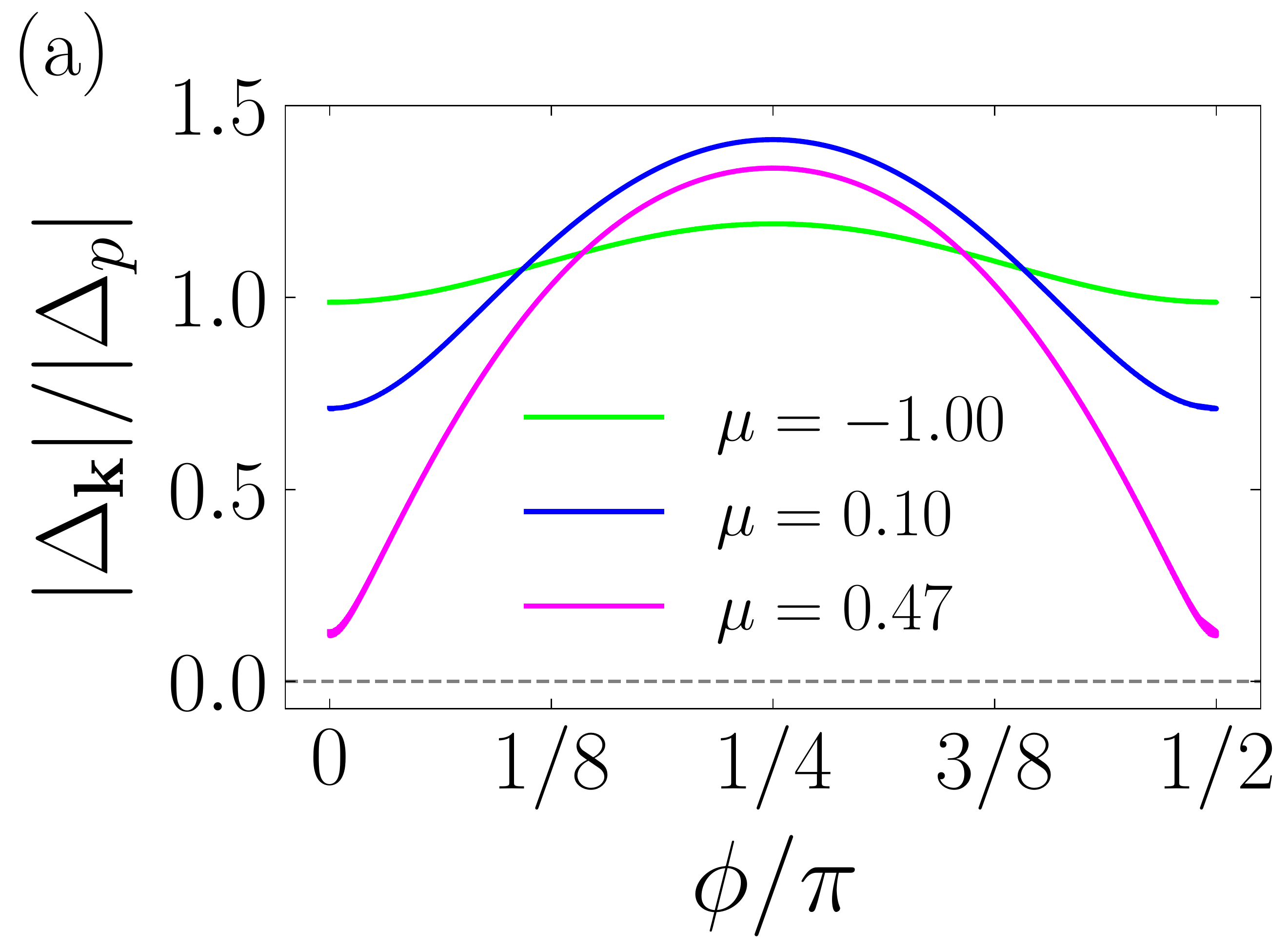}
	\end{minipage}
	\begin{minipage}[t]{0.49\columnwidth}
			\includegraphics[width=\textwidth]{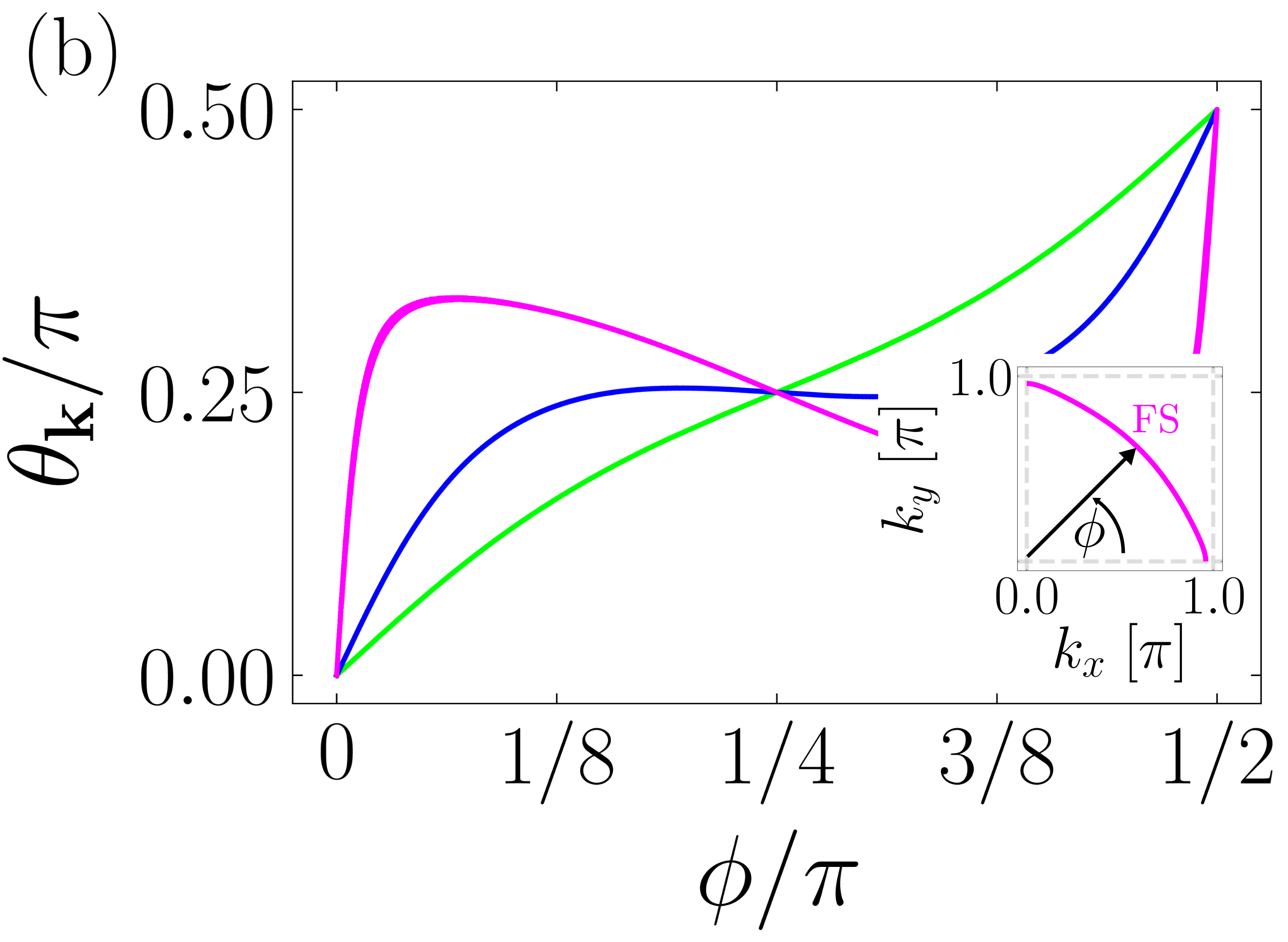}
	\end{minipage}
	\begin{minipage}[t]{0.49\columnwidth}
			\includegraphics[width=\textwidth]{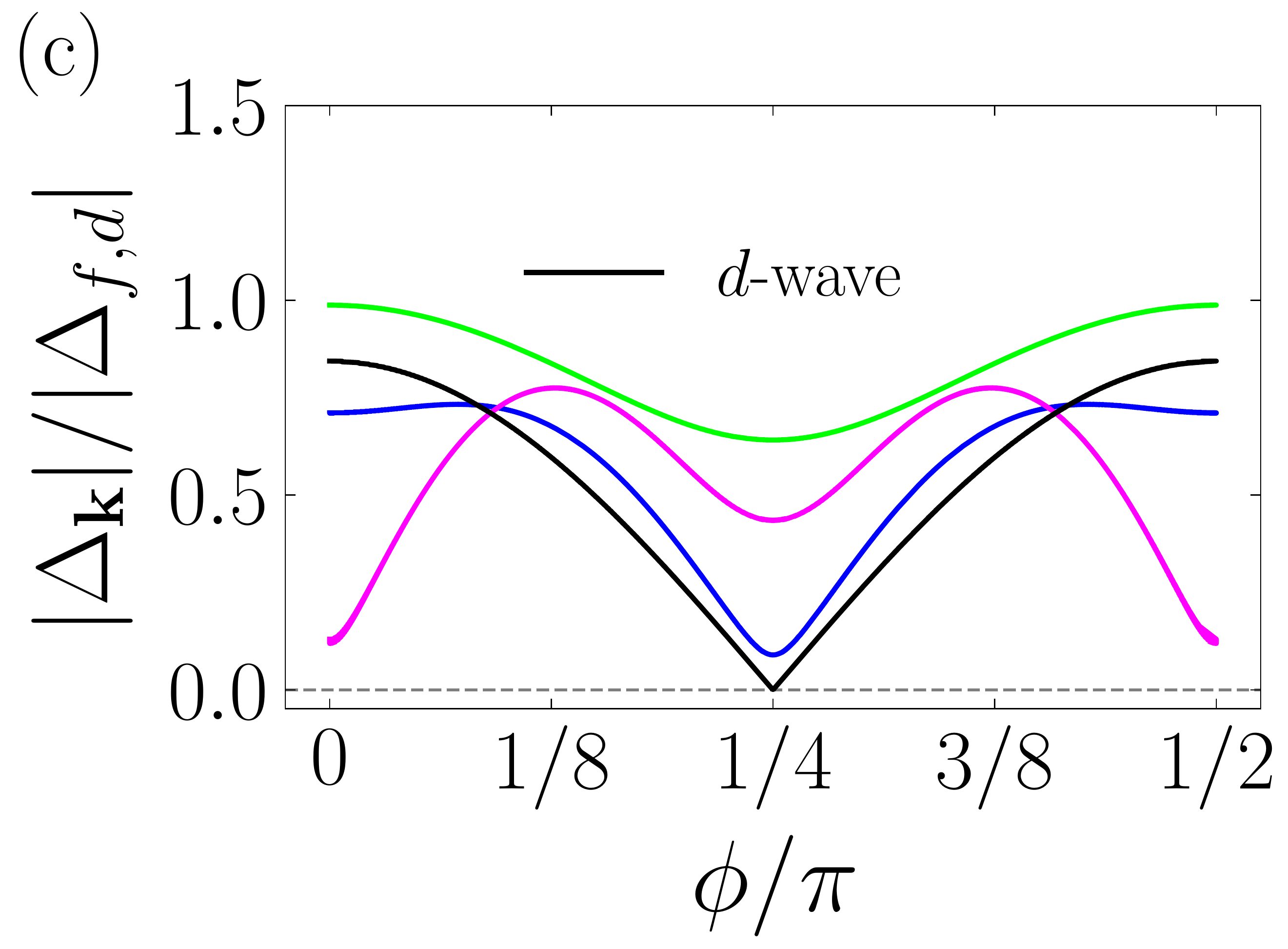}
	\end{minipage}
	\begin{minipage}[t]{0.49\columnwidth}
			\includegraphics[width=\textwidth]{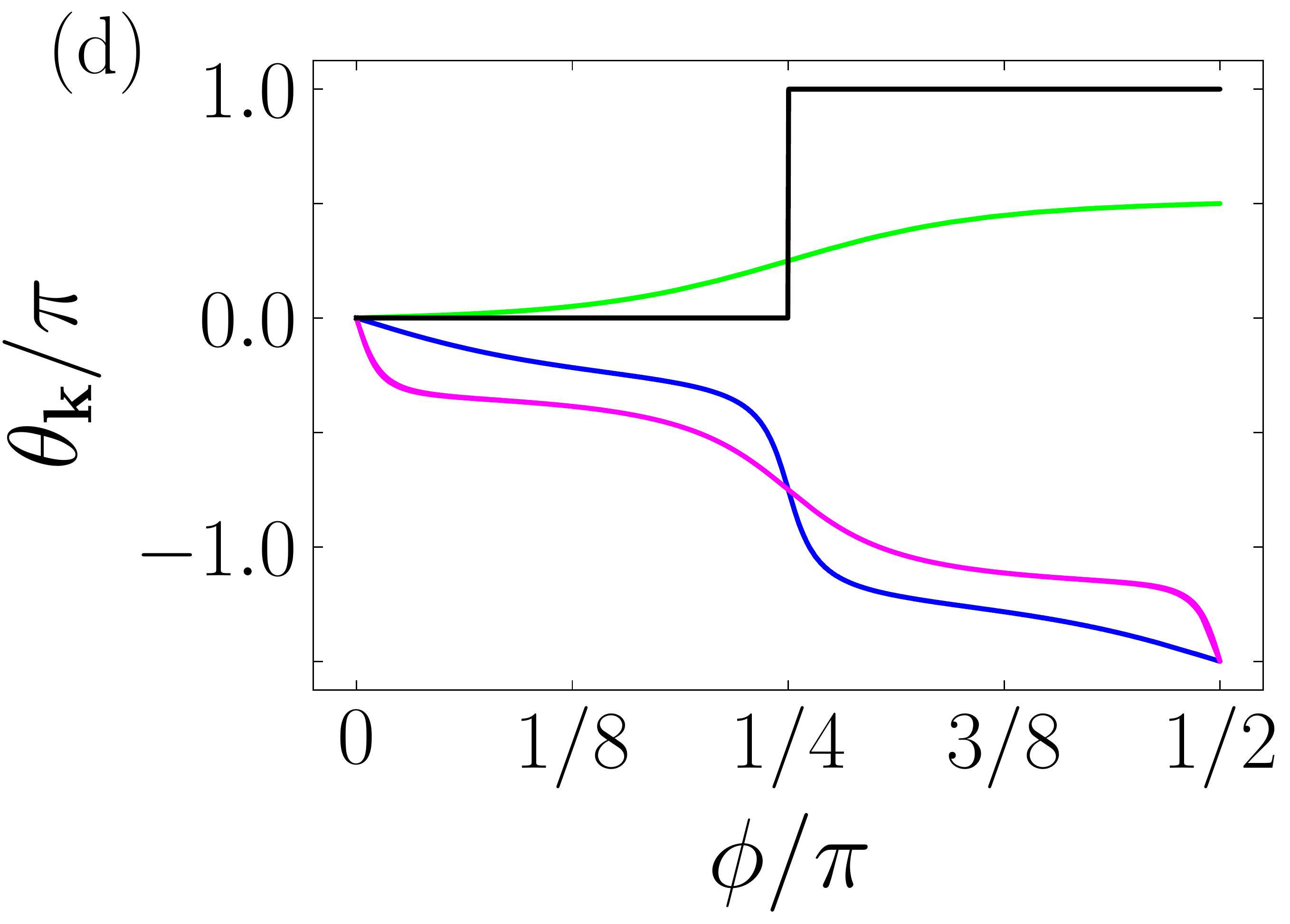}
	\end{minipage}
                
\caption{Gap anisotropy and phase winding for a $p$- and $f$-wave order parameter for different chemical potentials $\mu$, see Fig.~\ref{fig:nodes}. 
For better comparison, the gap magnitude is normalized with respect to $\Delta_p$ and $\Delta_f$, respectively.\\
}
\label{fig:anisotropy}
\end{figure}

Figures~\ref{fig:anisotropy} a) and b) show the gap function magnitude for the chiral $ p$-wave state and the phase $\theta_{\mathbf k}$ for the state with positive chirality along the normal-state Fermi surface for different band fillings. While the gap function has no nodes on the Fermi surface,  it shows strong minima for fillings, where the Fermi surface comes close to the van Hove points at $ (\pm \pi,0) $ and $( 0, \pm \pi )$, here for $ \mu = 0.47 $. Lowering the band filling leads to a more and more isotropic gap. For all band fillings shown the phase adds up to a total winding of $ 2 \pi $ and the Chern number is +1 for our choice of chirality. Interestingly, the angle dependence of the phase is non-monotonic for band fillings for which the Fermi surface crosses the lines connecting the van Hove points.

%

Figures~\ref{fig:anisotropy} c) and d) display the gap anisotropy and phase for the chiral $f$-wave state.  Again, we find a pronounced dip in the gap magnitude when the Fermi surface approaches the van Hove points ($\mu = 0.47 $). In addition, however, even more striking near nodes appear when the Fermi surface approaches the gap zeros at $(\pi/2, \pi/2)$. Further, we see how the phase now changes depending on whether the FS encompasses these points or not. If it does, the Chern number is $-3$, as one would expect from a chiral $f$-wave state in the isotropic case, while it is $+1$ otherwise.

Having both components $\Delta_f$ and $\Delta_p$, the
zeros corresponding to Eq.~\eqref{intra} can be shifted, while the zeros of Eq.~\eqref{inter} remain fixed at the BZ boundary. The former zeros
are located at the intersection 
of four lines satisfying the equations
\begin{eqnarray}
	0 &=& \Delta_p + \Delta_f \cos k_x, \label{z-1} \\
	0 &=& \Delta_p + \Delta_f \cos k_y , \label{z-2} 
\end{eqnarray}
such that the position along the $[11]$ direction depends on the ratio of $\Delta_p$ to $\Delta_f$. Note, however, that the winding numbers associated with the zeros remain. 
For $\Delta_p=\Delta_f$, the gap function develops zeros along the BZ boundary or main axes allowing for the winding numbers to change. Depending on the Fermi surface topology, this is accompanied with a topological transition in the superconducting state.

Before continuing with the analysis of the topological transition, we want to emphasize that the given discussion of the singularities and 
their topological properties can be applied to other systems as well. As an example, 
in App.~\ref{sec:hexagonal}, we comment on the situation of lattices with a three-fold rotation axis.


\subsection{Topological transition} 

In the following we will be interested in the behavior of superconductors when the zeros described by Eq.(\ref{z-1},\ref{z-2}) for $ | \Delta_p| <  |\Delta_f| $ lie close to the Fermi surface
and give rise to a near nodal spectrum. Therefore, it is important to know whether a near nodal situation is stable, if we give the positions of the zeros some flexibility combining $p$- and $f$-wave components. 
Within a weak-coupling picture one might expect that the order parameter would
change in a way as to avoid the zero near the Fermi surface, since it diminishes the condensation energy. 

Here, we analyze this question using a generalized Ginzburg-Landau theory for the two order parameters, whose coefficients are derived from 
the microscopic weak-coupling theory assuming a variation of the chemical potential to shift the Fermi surface as could be done by doping or
applying uniform pressure. The zero passing through the Fermi surface corresponds to a topological phase transition changing the Chern number. 
\begin{figure}[t!]
 \centering
\begin{minipage}[t]{0.49\columnwidth}
		\includegraphics[width=\textwidth]{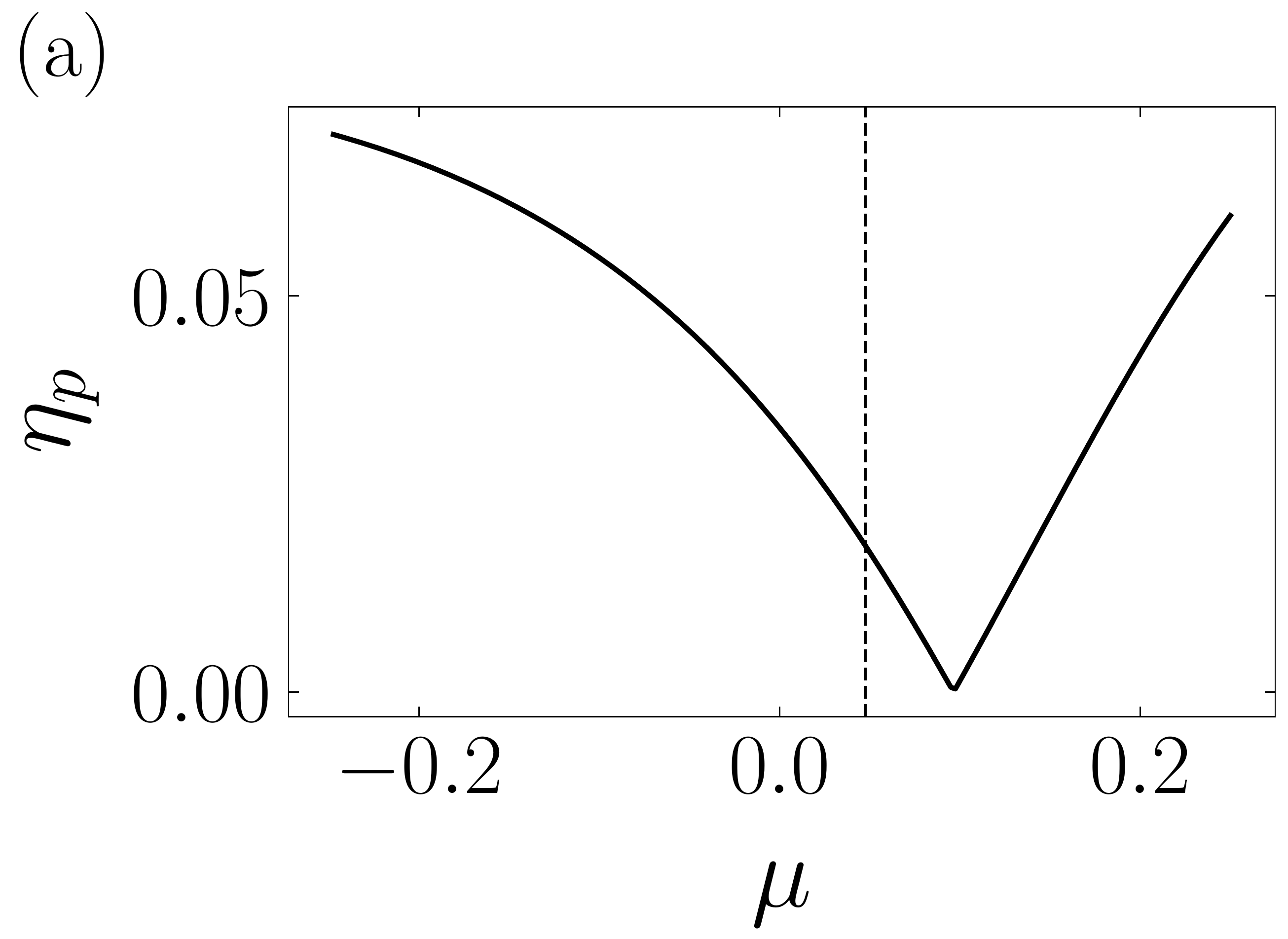}
\end{minipage}
\begin{minipage}[t]{0.49\columnwidth}
		\includegraphics[width=\textwidth]{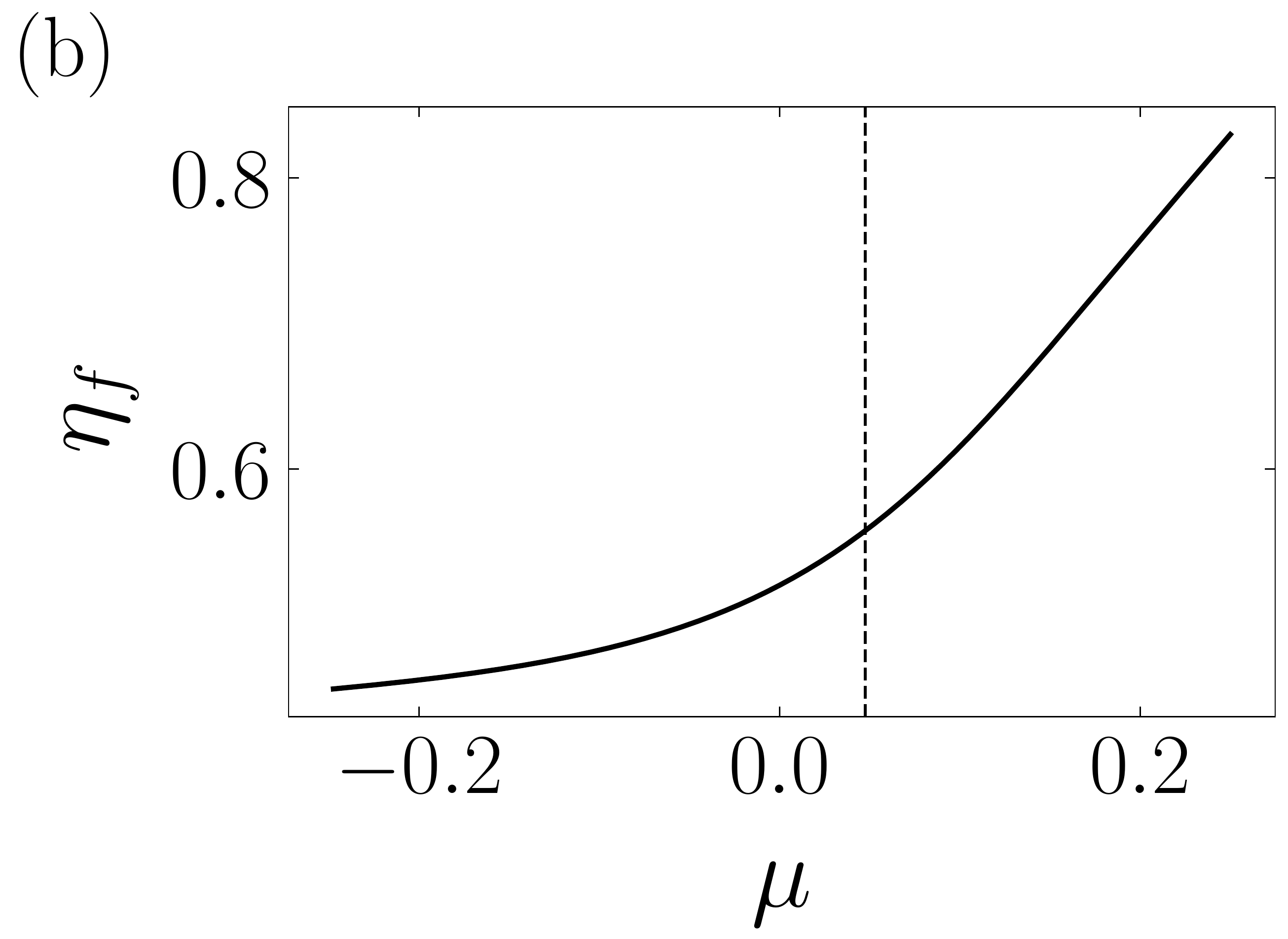}
\end{minipage}
\begin{minipage}[t]{0.49\columnwidth}
		\includegraphics[width=\textwidth]{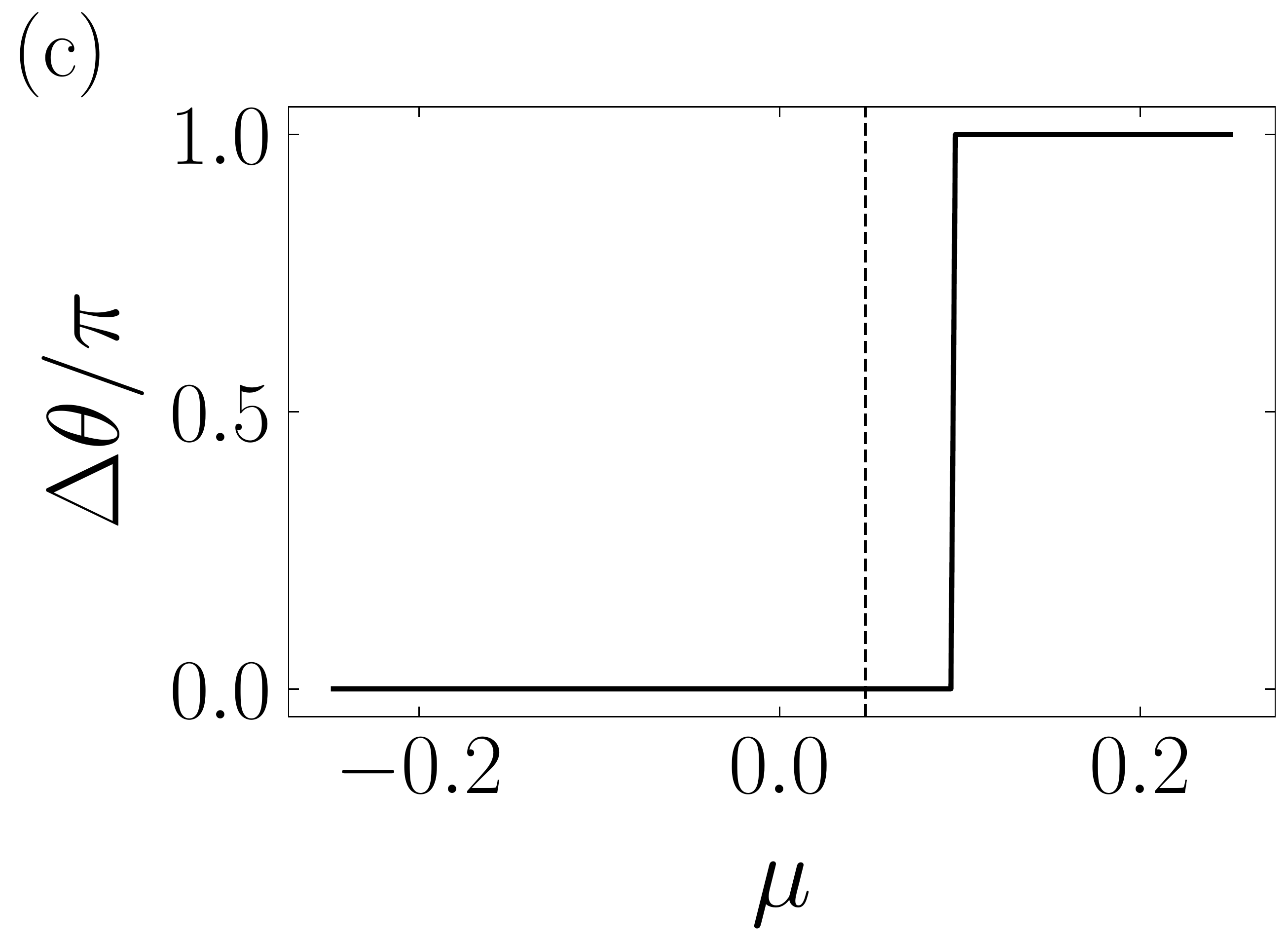}
\end{minipage}
\begin{minipage}[t]{0.49\columnwidth}
		\includegraphics[width=\textwidth]{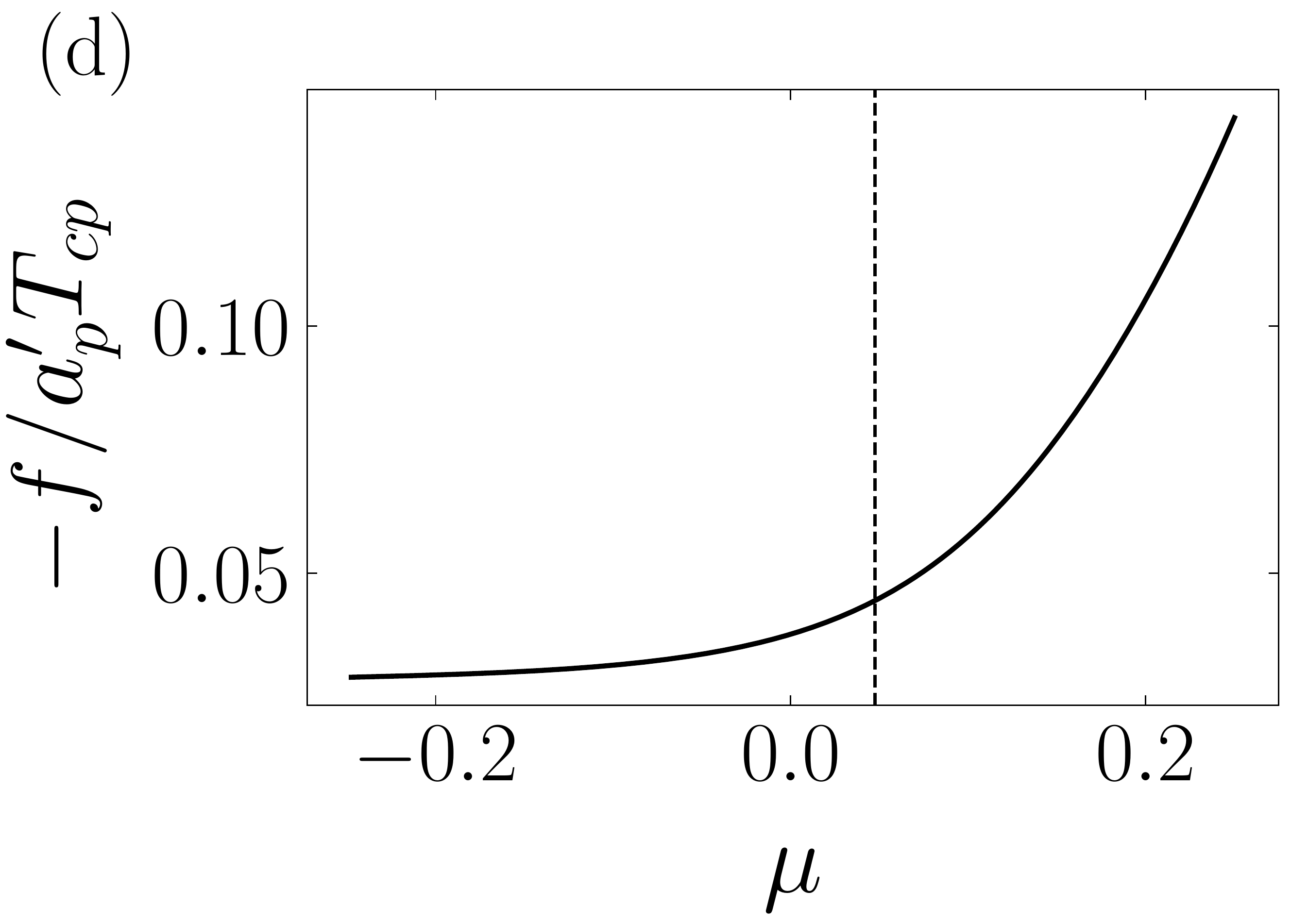}
\end{minipage}
\caption{
The order parameter components and free energy $f$ as a function of 
the chemical potential $\mu$ derived from the GL theory. The vertical 
dashed line illustrates the point, where the Chern number changes 
from $1$ to $-3$.
(a) The magnitude of the $p$-wave, $\eta_p$, in the uniform state.
(b) The magnitude of the $f$-wave, $\eta_f$, in the uniform state.  
(c) The phase difference between the $p$-wave and $f$-wave state, $\Delta\theta$. 
(d) The free energy normalized by the parameters $a'_p T_{cp}$.
}
\label{fig0.1}
\end{figure} 
The order parameter of the superconducting state 
for both $p$- and $f$-wave have two complex components and belong to the $ \Gamma_5^- $ representation of the tetragonal point group, 
\begin{align}
	\displaystyle \boldsymbol{\eta}_p &= (\eta_{p_{x}}, \eta_{p_{y}}), \nonumber \\[1mm]
	\boldsymbol{\eta}_f &= (\eta_{f_{x}}, \eta_{f_{y}}).
\end{align}
The Ginzburg-Landau free energy density for a simple two-component order parameter belonging to  $ \Gamma_5^- $  is given by \cite{sigrist1991phenomenological}
\begin{align}
	\displaystyle f &= a' (T - T_c) |\boldsymbol{\eta}|^2  + b_{1} |\boldsymbol{\eta}|^4 , \nonumber \\[1mm]
	&+ \frac{b_{2}}{2} (\eta_{x}^{\ast 2} \eta_{y}^{2} +  \eta_{x}^{2} \eta_{y}^{\ast 2}) + b_{3} |\eta_{x}|^2 |\eta_{y}|^2 ,
\end{align}
where we assume homogeneity and ignore gradient terms. 
Since both order parameters, $ \boldsymbol{\eta}_p $ and $ \boldsymbol{\eta}_f $ belong to $\Gamma_{5}^{-}$, they couple in a complex way. 
To simplify the problem, we assume for both order parameters a chiral state and make the ansatz 
\begin{align}
	\displaystyle \boldsymbol{\eta}_p &= \eta_p (1, i) e^{i\theta_p}, \nonumber \\[1mm]
	\boldsymbol{\eta}_f &= \eta_f (1, i) e^{i\theta_f}.
\end{align}
leading to the free energy expansion,
\begin{align}
	\displaystyle f &= 2 a'_p (T-T_{cp}) \eta_p^2 + 2 a'_f (T-T_{cf}) \eta_f^2 \nonumber \\[1mm]  
	&+ 4 \gamma \eta_p \eta_f \cos(\Delta\theta) + b_{p} \eta_p^4 + b_{f} \eta_f^4 \nonumber \\[1mm]  
	&+ 4 c_1 \eta_p^3 \eta_f \cos(\Delta\theta) + 4 c_2 \eta_p^2 \eta_f^2 \left[ 1 + \frac{1}{2} \cos(2\Delta\theta) \right] \nonumber \\[1mm] 
	&+ 4 c_3 \eta_p \eta_f^3 \cos(\Delta\theta), 
\end{align}
where $\Delta\theta = \theta_p - \theta_f$ and $b_{p,f} = 4b_{1p,f} - b_{2p,f} + b_{3p,f}$. 
The coefficients as a function of the chemical potential $ \mu $ are obtained through the expansion of
the self-consistent gap equation [Eq.~\eqref{eqn: sc-gap}]. 

Following the equilibrium state for varying $\mu$ we see in Fig.~\ref{fig0.1}
that no anomalous behavior appears when the zero crosses the Fermi energy. 
The critical $ \mu $ of the topological transition is indicated by the dashed line, where the Chern number changes
from 1 to - 3. Although both components change and shift the zeros slightly closer towards the $ \Gamma $ point,
they vary smoothly. This indicates that the zeros can be located arbitrarily close to the Fermi surface.

\section{Disorder and gap anisotropy}	 \label{sec: disorder}

Unconventional Cooper pairing is very susceptible to scattering even on non-magnetic impurities. Before analyzing the thermodynamic properties of the individual pairing possibilities, we thus consider
how disorder affects the two spin-triplet pairing states of $p$- and $f$-wave symmetry, introduced above, and combinations thereof.


\subsection{Theoretical framework}
We introduce disorder as identical impurity scatterers with concentration $c$. A single impurity at the
origin is described by
\begin{align}
	\mathcal{H}_{\text{imp}} = \sum_{\mathbf{k}, \mathbf{k'},s} U_{\mathbf{k}, \mathbf{k'}} c^{\dagger}_{\mathbf{k}s} c^{\phantom{\dag}}_{\mathbf{k'}s},
\end{align}
where $U_{\mathbf{k}, \mathbf{k'}}$  is the scattering matrix element. Since we consider strong scattering potentials, whose energy scale exceeds
the band width, we employ a T-matrix approach, which takes multiple scatterings at the same impurity into account \cite{hirschfeld1986resonant, hirschfeld1988consequences, balatsky2006impurity}.
The T-matrix is defined through the equation
\begin{align}
	T_{\mathbf{k}, \mathbf{k'}} (i \omega_n) 
	 &=  U_{\mathbf{k}, \mathbf{k'}} + \sum_{\mathbf{k''}} 
	U_{\mathbf{k}, \mathbf{k''}} G(\mathbf{k''}, i\omega_n) T_{\mathbf{k''}, \mathbf{k'}} (i \omega_n), 
\end{align}
where $G(\mathbf{k}, i\omega_n)$ is the (normal) electron Green's function. 
Note that we have omitted terms containing $\sum_{\mathbf{k}} F(\mathbf{k}, i\omega_n)$, where $F(\mathbf{k}, i\omega_n)$ is the anomalous Green's function, 
since these terms vanish for non-$s$-wave gap functions \cite{hirschfeld1988consequences, keller1988free, nomura2005theory}. 
Consequently, there are no off-diagonal entries in the T-matrix expansion, which would renormalize the gap directly. 
In the following, we consider isotropic, $s$-wave scattering for a point-like impurity potential, removing the momentum dependence from the scattering matrix elements $U_{\mathbf{k}, \mathbf{k'}} = U$. Thus, the T-matrix is a scalar in momentum space as well, 
\begin{align}
	T_{\mathbf{k}, \mathbf{k'}} (i \omega_n) = T(i \omega_n). \label{eqn: iso-T}
\end{align}
Restricting our investigation to small impurity concentrations $c$, we can safely neglect impurity interference effects coming from multiple scatterings on different impurities.
Hence, the self-energy is proportional to $c$, 
\begin{align}
	\Sigma(i \omega_n) = c T(i \omega_n) \label{eqn: eq3-Sigma}
\end{align}
and can be absorbed into a renormalized Matsubara frequency
\begin{align}
	  i\tilde{\omega}_n = i\omega_n - \Sigma(i\omega_n). \label{eqn: eq3-omeg}
\end{align}
The normal and anomalous Green's functions and the self-energy are then related through Gorkov's equations~\cite{mineev1999introduction, nomura2005theory}. 
Assuming a flat density of states over an energy scale larger than the gap size, we obtain the following simplified form for the normal and anomalous Greens functions, 
\begin{align}
	G(\mathbf{k}, i\omega_n) &= -\frac{i\tilde{\omega}_n + \xi_{\mathbf{k}}}{\tilde{\omega}_n^2 + \xi_{\mathbf{k}}^2 + |\Delta_{\mathbf{k}}|^2}, 
	\label{eqn: eq4} \\[1mm]
	  F(\mathbf{k}, i\omega_n) &= \frac{\Delta_{\mathbf{k}}}{\tilde{\omega}_n^2 + \xi_{\mathbf{k}}^2 + |\Delta_{\mathbf{k}}|^2}, \label{eqn: eq7} \\[1mm] 
	  F^{\dagger}(\mathbf{k}, i\omega_n) &= \frac{\Delta_{\mathbf{k}}^{\ast}}{\tilde{\omega}_n^2 + \xi_{\mathbf{k}}^2 + |\Delta_{\mathbf{k}}|^2}, \label{eqn: eq8}
\end{align}
where Eq.(\ref{eqn: eq3-omeg}) has to be determined self-consistently. 
Note that the assumption of a flat density of states is generically not justified, but still affects results for the thermodynamic response considered below in
a negligible way \cite{monien1987resonant, nomura2005theory}.  
Using the renormalized Matsubara frequencies in the self-consistent 
gap equation [Eq.~\eqref{eqn: sc-gap}], we can examine the influence of disorder
on the various spin-triplet pairing states.


\subsection{Influence on gap anisotropy}

If we consider the two spin-triplet pairing states separately, disorder treated within the above formalism merely leads
to a gradual shape-conserving suppression of the gap magnitude ($ \Delta_p $ or $ \Delta_f $) with increasing $c$. 
However, we can study the influence of disorder on the gap anisotropy by studying a combination of $p$- and $f$-wave 
state. For this purpose, we treat simultaneously NN- and NNN-pairing interactions self-consistently, which results 
in a mutual interaction between the two pairing channels.  
For the chemical potential, we choose $ \mu = 0.1 $, such that the Fermi surface passes rather close to the $f$-wave zeros at $\mathbf{k} = (\pm \pi/2, \pm \pi/2)$. 
\begin{figure}[t!]
 \centering
\begin{minipage}[t]{0.90\columnwidth}
		\includegraphics[width=\textwidth]{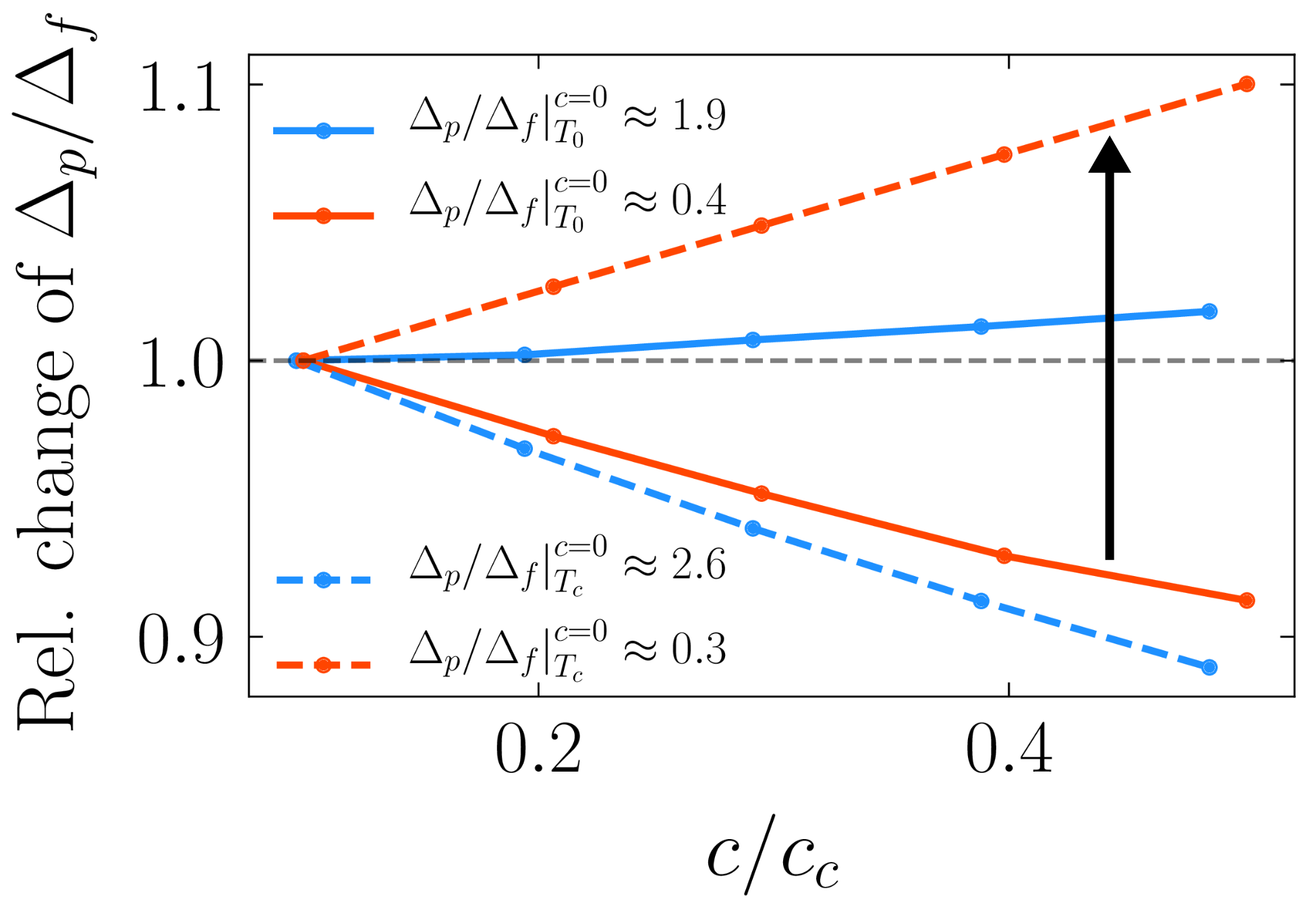}
\end{minipage}
\caption{
The relative change of the gap ratio $\Delta_p / \Delta_f$ at $\mu = 0.1$ as 
a function of the impurity concentration $c$.  
This change is shown for two different combinations for
the temperatures $ T_c $ and $ T_0 = T_c / 50 $, where $ T_c $ depends on the impurity concentration. 
The impurity concentration $c$ is
normalized to the critical value $c_c$. 
}
\label{figGapRatios}
\end{figure}

In Fig.~\ref{figGapRatios}, we show the relative change of the ratio $ \Delta_p / \Delta_f $ for two sets of pairing interactions.
Both the temperature and the impurity concentration change this ratio. For a dominant $p$-wave component,
we find a decrease of the ratio with temperature from $  \Delta_p / \Delta_f  \approx 2.6 $ at $ T_c $ to $ 1.9 $ at $ T_0 = T_c/50$ in the clean limit.
Interestingly, the ratio decreases with disorder at $ T_c $ while this trend is reversed at low temperature. 
At low temperature, the change can be understood from the difference in the coherence length of the two pairing states, whereby the
one with the shorter coherence length is at an advantage. The coherence length can be defined as
\begin{align}
	\xi^2 =  \frac{\sum_{\mathbf{k}} \Big | \nabla_{k} \frac{\Delta_{\mathbf{k}} }{E_{\mathbf{k}}}\Big |^2}{\sum_{\mathbf{k}} \Big | \frac{\Delta_{\mathbf{k}}}{E_{\mathbf{k}}}\Big |^2 } .
\end{align}  
for both states separately. For dominant $ p$-wave pairing, a ratio of $ \xi_p/\xi_f \approx 0.24 $ at $T_0$ thus implies that the $f$-wave component is more fragile.
At $T_c$, however, the coherence length of the (total) order parameter diverges. Being closer to its pure $T_c$, the dominant channel has a longer coherence length and is thus 
more fragile.
This behavior is corroborated by the opposite case observed for a stronger $ f $-wave pairing interaction, where at $ T_c $ disorder favors the weaker state while at $ T_0 $ the $p$-wave state is more
quickly suppressed by impurities with $ \xi_p/\xi_f \approx 1.25 $. 
Note, finally, that all these findings apply also for combination with a relative phase of $ \pi $ between the two components. In other words, the influence of impurity scattering is generally \emph{not} a reduction in anisotropy, but rather a suppression of the component with larger coherence length.

\section{Thermodynamic properties}  \label{sec: td}
We now turn to the (low-temperature) thermodynamic properties of the various states discussed above. For comparison to 
a truly nodal state, we additionally study a $ d_{x^2-y^2} $-wave pairing state. While the clean-limit behavior at $T\ll T_c$ is entirely 
dominated by the presence or absence of nodes, we show in the following how disorder complicates the picture. Finally, as the real
situation is generically not a pure $p$-wave or pure $f$-wave order parameter, we study the influence of the gap anisotropy given by changing the 
ratio $\Delta_p/ \Delta_f$.


\subsection{Pure pairing channels}
\begin{figure}[t!]
 \centering
    \includegraphics[width=0.90\columnwidth]{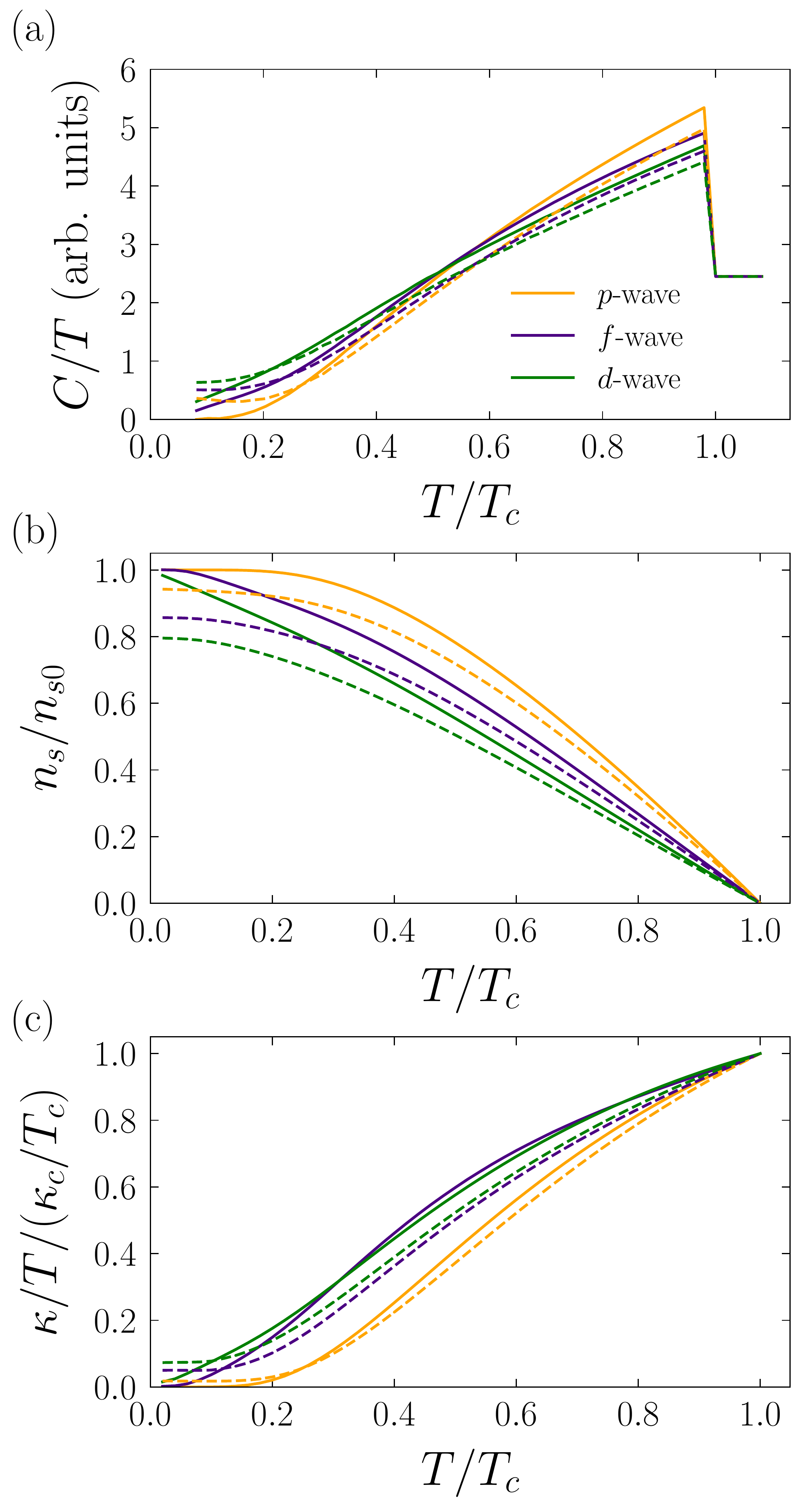}
\caption{
Summary of thermodynamic properties: (a) specific heat, (b) superfluid density, and (c) thermal conductivity 
as a function of temperature $T$ for a $p$-, $f$- and $d$-wave state at $\mu = 0.1$. 
The solid (dashed) line illustrates the results obtained for zero (finite) disorder. 
In the case of finite disorder, the impurity concentration was $c/c_c \approx 0.12$.
The superfluid density is normalized to the zero-temperature, 
zero-impurity constant, $n_{s0}$.
In the case of $\kappa$, a finite impurity concentration is necessary. Therefore, the solid (dashed) lines show results
for a concentration of $c/c_c \approx 0.05$ ($c/c_c \approx 0.28$).\\
}
\label{fig:allTD}
\end{figure}

First, we study the three pairing states independently including disorder by means of the Greens function formalism 
introduced above and linear response theory \cite{luttinger1960ground, langer1962js, ambegaokar1964theory, 
luttinger1964theory, schrieffer2018theory, keller1988free, lee1997unusual}. 
For this purpose, we calculate the specific heat, the superfluid density, and the thermal conductivity. 
A short overview over their derivations is provided in App.~\ref{sec: appendixB}.  
Our results are shown in Figs.~\ref{fig:allTD} (a)-(c).

Figure~\ref{fig:allTD} (a) shows the specific heat divided by the temperature, $C/T$. In the clean limit, 
we observe that both $f$- and $d$-wave exhibit a linear temperature dependence down to low temperatures. 
While the linear dependence goes all the way to $T=0$ for the $d$-wave case, the $f$-wave state
with no true node has a downwards bending at low temperatures. 
In the same temperature range, the $p$-wave state clearly shows the exponential suppression of the specific heat at low temperature. 
For all three states, there is no residual value of $ C/T $ at zero temperature due to the vanishing density of 
states in the clean system. This changes with disorder, which leads to a non-vanishing $ C/T $ 
at $ T=0 $ all three states. Importantly, impurities lead to an even stronger qualitative resemblance of the $ f$-wave and the $d$-wave state. 
Note that for the calculations the same chemical potential was used for all three pairing states.

Turning to the superfluid density, Fig.~\ref{fig:allTD} (b), both spin-triplet states exhibit 
an exponential saturation of the superfluid density at low temperatures due to their fully gapped excitation spectrum.
The nodal $d$-wave case, on the other hand, shows the expected linear temperature dependence in the clean limit.
Once disorder is included, however, the $f$-wave and $d$-wave states can hardly be distinguished anymore.
Both curves have approximatively the same slope and only a different residual value at zero temperature.

Finally, the thermal conductivity, shown in Fig.~\ref{fig:allTD} (c), yields a similar picture;  
In the clean limit, both the nodal spin-singlet and the very anisotropic $f$-wave state look much alike on
an overall level but have a clear qualitative difference at very low temperature, where the former vanishes linearly
in temperature and the latter has an exponential suppressing indicating a tiny gap. The suppression
is naturally more pronounced for the large gap $ p $-wave state.
Much like the above discussed specific heat and superfluid density, disorder renders the $d$- and $f$-wave states practically indistinguishable 
with only different residual values at $ T= 0 $. Again, the $ p$-wave state shows a much stronger suppression
and lower residual conductivity.
 
 \begin{figure}[t!]
 \centering
		\includegraphics[width=0.90\columnwidth]{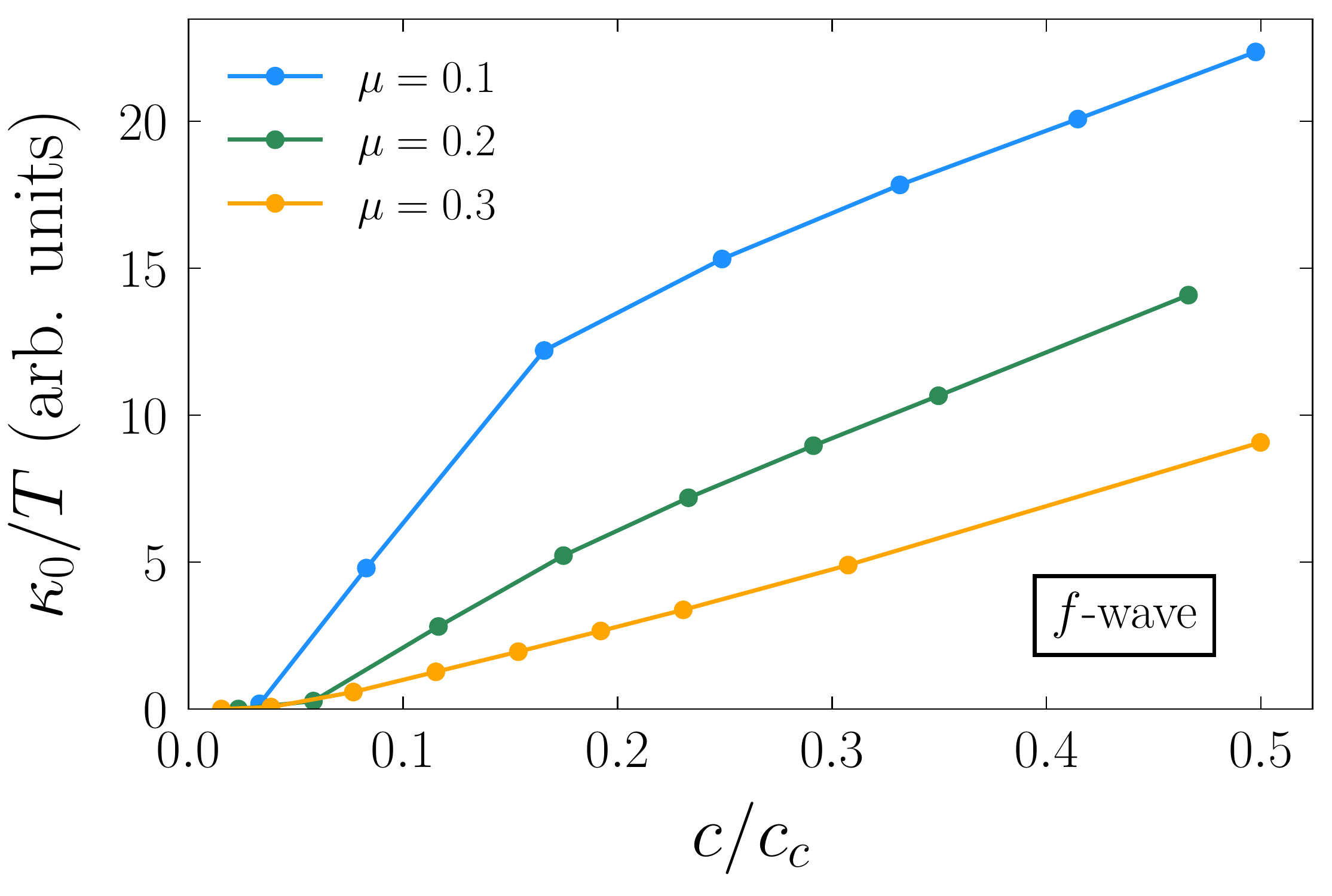}
\caption{
The residual thermal conductivity $\kappa_0 = \kappa(T=0)$ of the $f$-wave state as a function of 
impurity concentration $c$ ($c_c$ is the critical concentration) at 
different values of the chemical potential $\mu$. Note that we have 
used a smaller gap for the computation of this figure compared to the 
previous ones. This explains the different magnitude of $\kappa_0/T$.
}
\label{figMuChange}
\end{figure}

The residual value of the thermal conductance, in particular its saturation for finite impurity density, is often used as a 
characterizing feature of symmetry-imposed nodes, such as in the case of a $d$-wave state. Indeed, we see in Fig.~\ref{figMuChange},
that the residual value for the system with near nodes always vanishes in the limit of low impurity concentrations. However, for deep enough nodes, the 
concentration necessary to see this trend can be very low.

\begin{figure}[t!]
 \centering
\begin{minipage}[t]{0.49\columnwidth}
		\includegraphics[width=\textwidth]{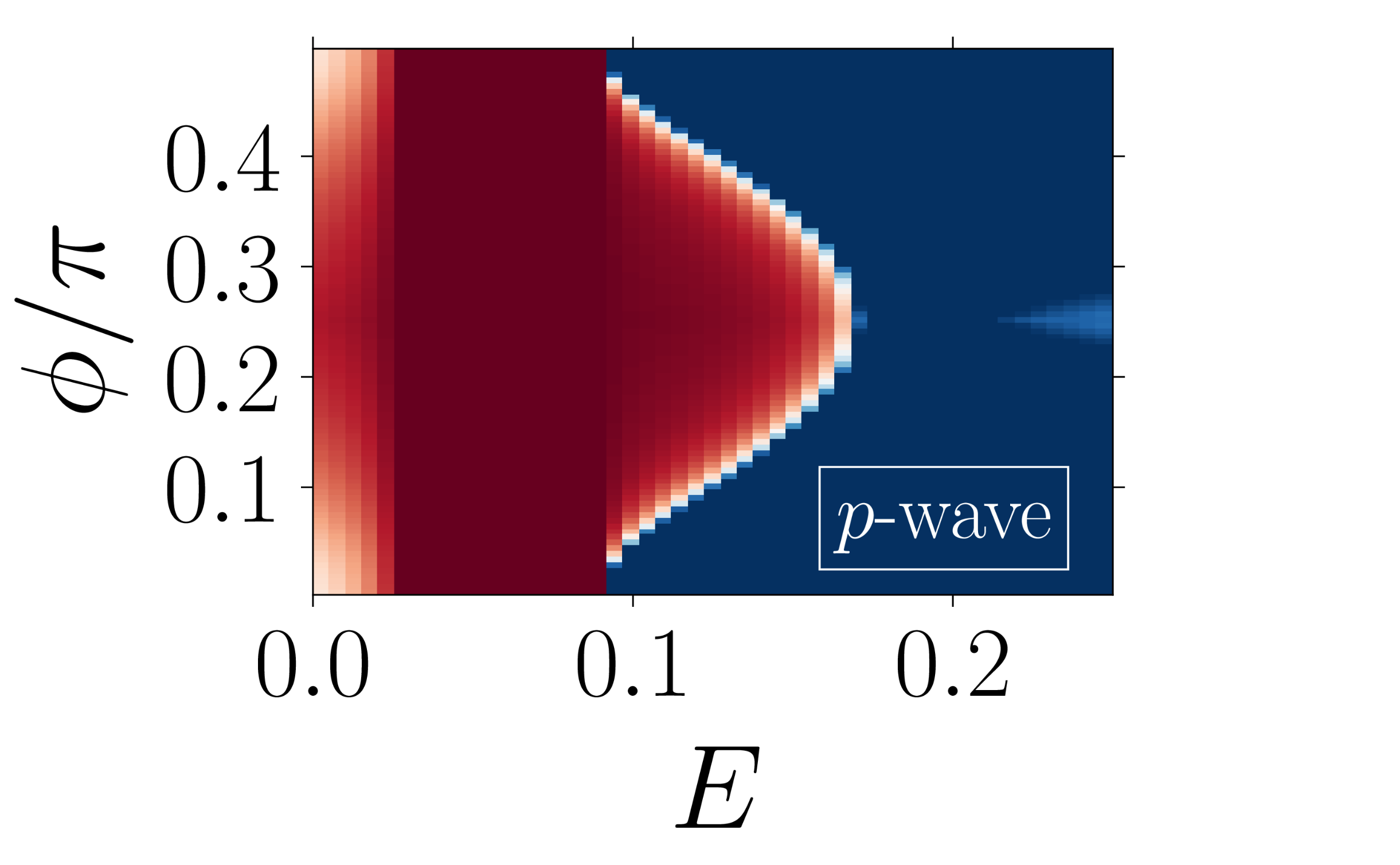}
\end{minipage}
\hspace{-0.8cm}
\begin{minipage}[t]{0.49\columnwidth}
		\includegraphics[width=\textwidth]{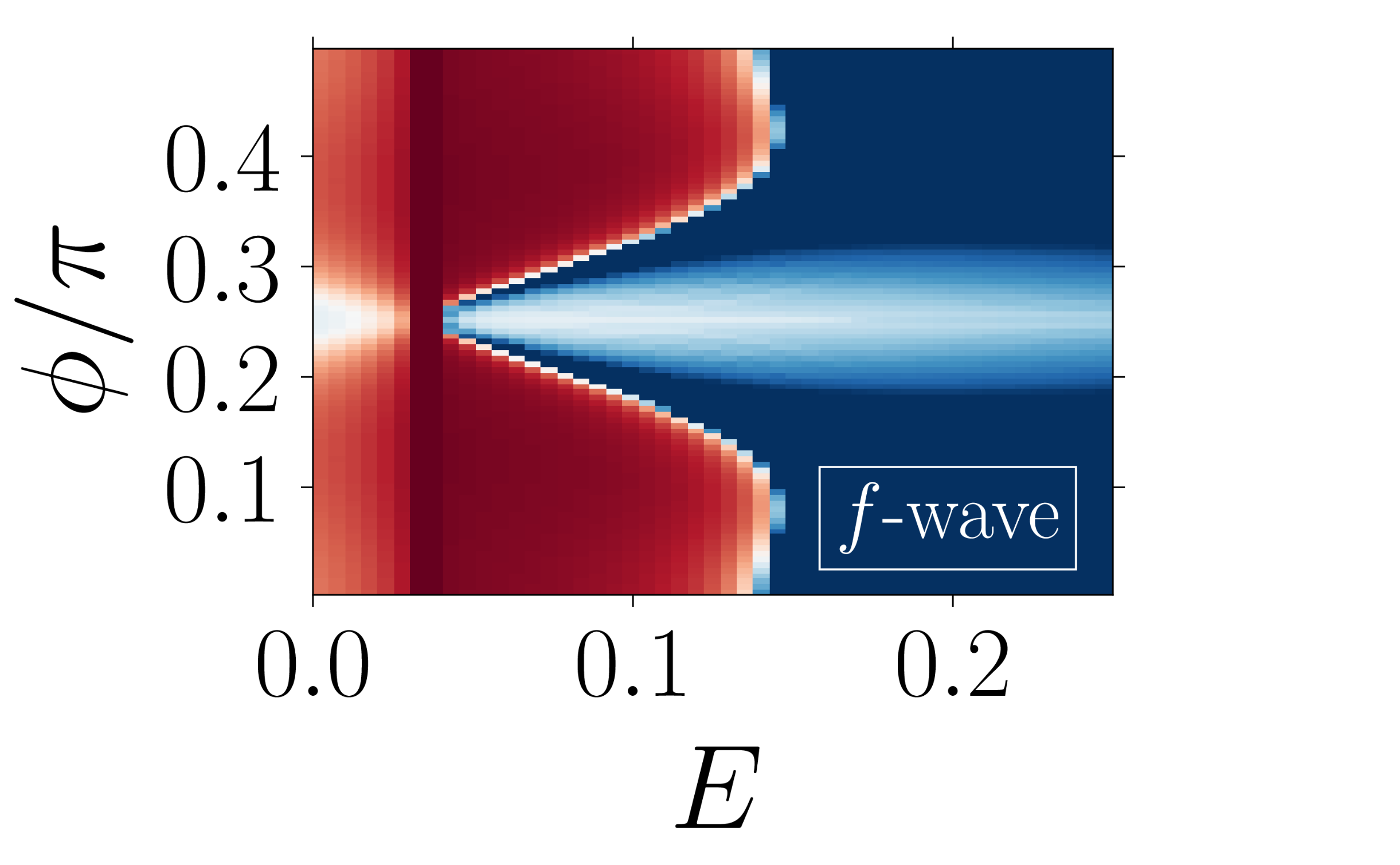}
\end{minipage}
\begin{minipage}[t]{0.49\columnwidth}
		\includegraphics[width=\textwidth]{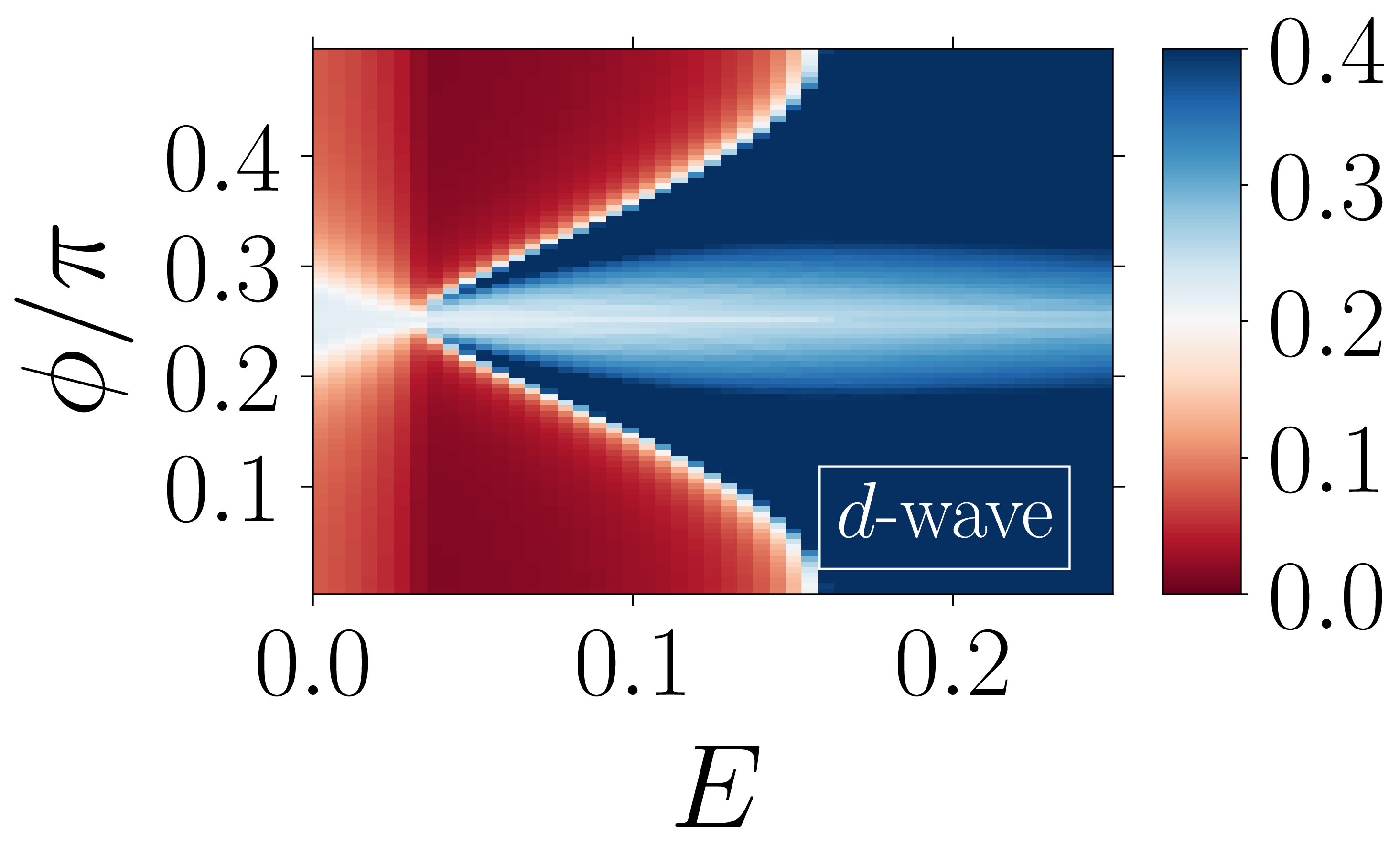}
\end{minipage}
\hspace{0.3cm}
\caption{
The density of states as a function of the angle $\phi$ 
and quasiparticle energy $E$ calculated for the $p$-,$f$- 
and $d$-wave state at the impurity concentration $c/c_c \approx 0.18$.
}
\label{fig:dos}
\end{figure}  

We can shed some light into the general trends shown in Figs.~\ref{fig:allTD} and \ref{figMuChange} 
by looking at the angle-resolved density of states. Figure~\ref{fig:dos} shows the 
radially-integrated density of states for a moderate impurity concentration of $c/ c_c = 0.18 $ restricted to a quarter of the BZ, $ \phi \in [0,\pi/2] $. 
For all three states, there is a clear, slightly softened gap with excitations above it. 
The $d$-wave and $f$-wave states have very similar density of states due to their nodal or near-nodal structure, with low-energy states induced by the impurity scattering well visible. 
The dominant feature appears around the nodal direction ($\phi = \pi/4 $) with a peak in the
density of low-energy states whose weight has been shifted from the higher-energy region. 
The region of the nodes has the highest Fermi velocity, making states in this region especially important for heat transport. 
It is the relocation of states that is crucially affecting the
low-temperature behavior and the two pairing states' resemblance observed in the thermodynamic properties.
In the case of $p$-wave pairing, the impurities only give rise to subgap states at low energy, with a broad accumulation, where the Fermi surface approaches the van Hove points ($ \phi =0 , \pi/2 $).
The low-temperature behavior is thus still close to that of a fully gapped system with only little residual density of states at $T=0$.
Finally, note again how Fig.~\ref{figMuChange} shows that the ''nodal'' behavior of the $ f$-wave state 
depends crucially on how close the gap zero is to the Fermi surface. 
\begin{figure}[t!]
 \centering
		\includegraphics[width=0.90\columnwidth]{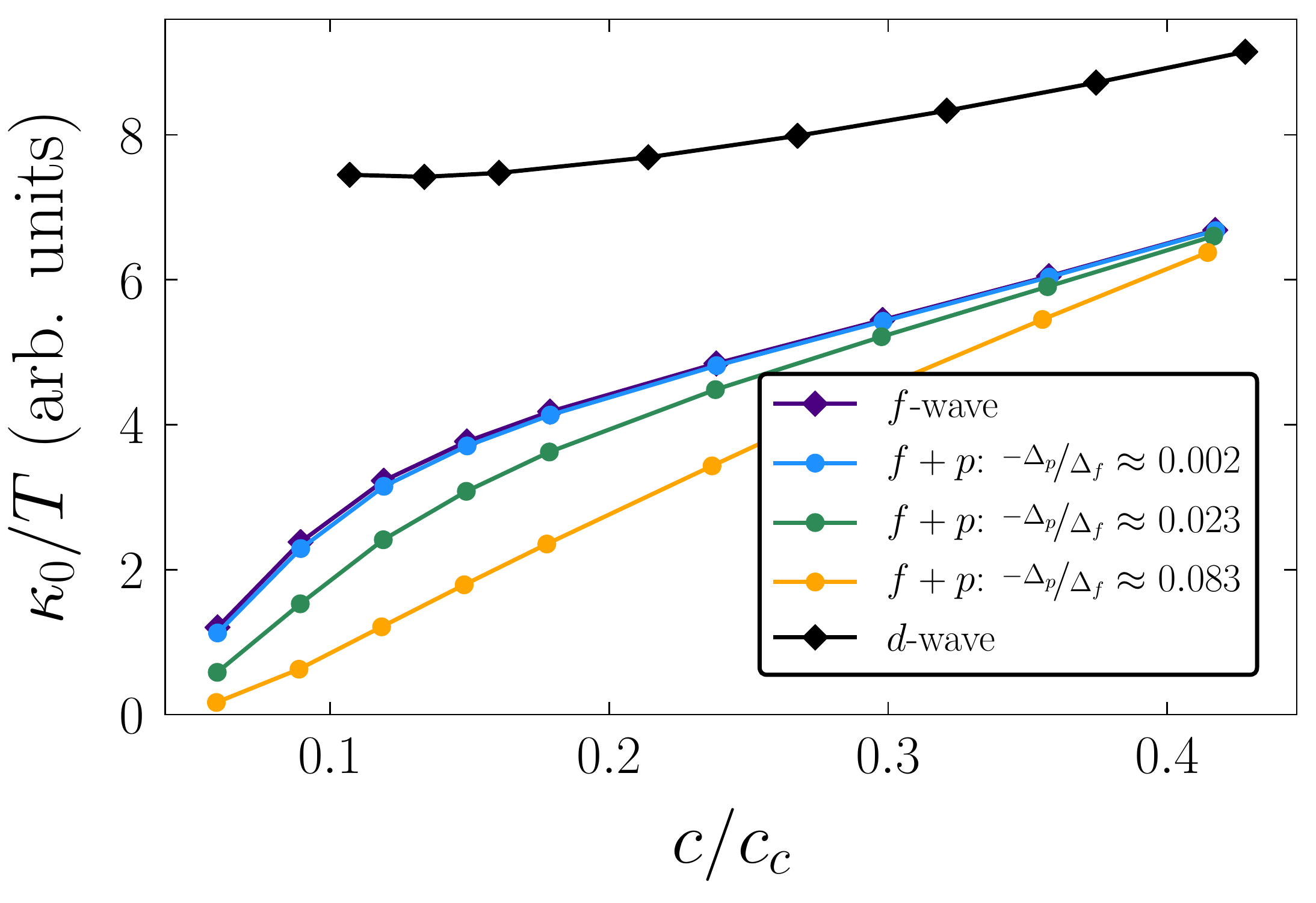}
\caption{
The residual thermal conductivity 
as a function of impurity concentration at $\mu = 0.1$ 
for different superpositions of $p$-wave and $f$-wave.}
\label{fig5}
\end{figure}



\subsection{Tuning the spin-triplet anisotropy strength}
While we have restricted our discussion of thermodynamic properties to pure $p$- or $f$-wave states so far,
we expect a superposition of the two in the general case. The effect of impurities in this case is then twofold: First, it 
changes the ratio of the two channels, thus changing the anisotropy. Second, the impurities lead to low-energy density of
states, which depends on the anisotropy of the gap in the excitation spectrum.
Therefore, we finish our discussion of thermodynamic properties with the case of different superpositions.
In the following, we assume a relative phase of $\pi$ between the two components,
such that the zeros of the gap are moved from $\mathbf{k} = (\pm \pi/2, \pm \pi/2)$ towards the BZ center 
and are always enclosed by the Fermi surface ($ \mu = 0.1 $). Under these conditions the increase of $|\Delta_p|$ relative 
to $|\Delta_f|$ leads to a more isotropic gap, since the gap zero moves away from the FS. 

In order to illustrate the effect of disorder on the superposition, we investigate the heat conductance
at very low temperature and follow the residual value, $ \kappa_0 / T $ for $ T \to 0 $. In Fig.~\ref{fig5}, we
display $\kappa_0 / T$ for different ratios $ \Delta_p / \Delta_f $ at $c=0 $ and compare it also with the nodal $ d $-wave gap. 
We observe that for the $d$-wave state, $ \kappa_0 /T $ varies little with $c$. On the other hand, for all
considered ratios $ \Delta_p / \Delta_f $ the values of $ \kappa_0 / T $ are suppressed to zero for $ c \to 0 $ and
approach each other with increasing $ c $. While these residual heat conductance behave similar to 
the $d$-wave state for larger $ c/c_c $, they always lie lower. As Fig.~\ref{figMuChange} shows the residual value
increases the closer the gap zeros come to the FS, when we compare the $f$-wave states for different
chemical potentials. From both Figs.~\ref{fig5} and  \ref{figMuChange} we see that the characteristic 
feature of downwards bending of $ \kappa_0 / T $ for $ c $ going towards zero is gradually disappearing 
when the FS is removed from the gap zero. This behavior is characteristic for 
near nodal states.

\section{Discussion and Conclusion} \label{sec: conclusion}

Going beyond a pure point group analysis of gap-function zeros for chiral superconducting state,
we have found a rich structure of gap singularities and associated winding numbers in the Brillouin zone.
Depending on the exact form of the order parameter and the position of the Fermi surface, this structure influences the 
topological invariant of the system, the Chern number, and can lead to a pronounced anisotropy in the gap function as defined 
along the normal state Fermi surface. 
This anisotropy in the excitation spectrum has clear signatures in the thermodynamic properties of the system, 
exemplified by the specific heat, superfluid density, and thermal transport in Fig.~\ref{fig:allTD}. For a Fermi surface close to the gap-function zeros,
the anisotropic chiral $f$-wave state becomes almost indistinguishable from a (nodal) $d$-wave state for even small disorder.

A generic order parameter is not a pure chiral $p$-wave or $f$-wave state, but a superposition of the two (and higher angular momentum channels).
This raises two important issues that we have addressed in this work: First, is a gap-function zero close to the Fermi surface stable, or will the system 
find a superposition of $p$-wave and $f$-wave states to avoid this situation? Employing a Ginzburg-Landau analysis of this situation, we have 
indeed found that the zeros are generally not avoided. Instead, we find a topological transition when moving the Fermi surface through the zeros, which is of second order, see Fig.~\ref{fig0.1}.

A second important issue concerns the influence of disorder on the gap anisotropy.
Given the basic notion that disorder leads to a more isotropic order parameter, it is crucial to study the evolution of anisotropy in the presence of disorder for a superposition of two unconventional states.
Interestingly, we find that the anisotropy can both increase 
and decrease when the impurity concentration is increased. 
We explain our unusual observation by first noting that both spin-triplet states are unconventional in nature and break time-reversal symmetry. 
Therefore, both gap functions are suppressed by disorder \cite{anderson1959theory}.
For substantially different gap contributions of the two states, 
the respective coherence lengths ($\xi \propto 1/|\Delta|^2$) differ as well and with it the two components' susceptibility to disorder.
The resulting stronger suppression of the pairing channel with longer coherence length can both increase and decrease the 
gap-function anisotropy.

To conclude, our work highlights the difficulty in interpreting thermodynamic measurements in case of chiral superconducting order 
parameters. Taken together with \cite{wang2019, kallin2018}, our findings show that thermodynamic measurements to distinguish a chiral
$f$-wave state from a $d$-wave state are not necessarily
conclusive even when considering different impurity concentrations. Such an analysis was done recently for Sr$_2$RuO$_4$~\cite{hassinger2017vertical} and 
motivated this work.
However, there is a strong dependence of the anisotropy of an $f$-wave state on the Fermi
surface position, in contrast to a $d$-wave state with point-group symmetry imposed nodes. 
Thus, Fermi surface engineering, as for example suggested by \textcite{hsu2016manipulating}, 
combined with thermodynamic measurements could provide a path forward.

\section*{Acknowledgements}
We are very grateful to Jose L. Lado, Jonathan Ruhman, and Catherine Kallin for many useful discussions. 
This work was financially supported by the Swiss National Science Foundation (SNSF) through Division II (No. 184739).

\appendix
\section{Singularities of gap functions due to lattice translational symmetries} \label{sec:singularities}

Using inversion in Eq.~\eqref{eq:trafos}, 
we generally find for an odd-parity order parameter ($\phi_{\rm I} = \pi$)
\begin{align}
	\Delta_{\mathbf{k}} = - \Delta_{-\mathbf{k}} = -\Delta_{-\mathbf{k} + \mathbf{G}}.
\end{align}
Together with Eq.~\eqref{eq:morezeros}, we identify 
\begin{align}
	\mathbf{k} = (0, 0),\ (\pi, 0),\ (0, \pi),\ (\pi, \pi), \label{inter}
\end{align}
and all equivalent points in the BZ as zeros of $\Delta_{\mathbf{k}}$.

Since the square lattice is bipartite, the NN (NNN) pairing states couple electrons on different (the same) sublattices. This leads to a distinguishing property
between inter- and intra-sublattice pairing states, namely
\begin{equation}
\Delta_{\mathbf{k}}^{\rm inter} = - \Delta_{\mathbf{k}+\mathbf{Q}}^{\rm inter} \label{Q-inter}
\end{equation}
and
\begin{equation}
	\Delta_{\mathbf{k}}^{\rm intra} = + \Delta_{\mathbf{k}+\mathbf{Q}}^{\rm intra}   \label{Q-intra}
\end{equation}
with $ \mathbf{Q} = (\pi, \pi) $. Note that this is a `reciprocal lattice vector' considering a folded BZ. For intra-sublattice states, we can use the relations
\begin{equation}
	\Delta_{\mathbf{k}}^{\rm intra} = - \Delta_{- \mathbf{k}}^{\rm intra} = - \Delta_{- \mathbf{k}+ \mathbf{Q}}^{\rm intra}
\end{equation}
to identify, through $ \mathbf{k} = - \mathbf{k}+ \mathbf{Q} $,  another set of points
\begin{equation}
	\mathbf{k} = (\pm \pi/2 , \pm \pi/2 ) \label{intra}
\end{equation}
with $\Delta_{\mathbf{k}}^{\rm intra}=0 $. $\Delta_{\mathbf{k}}^{\rm inter}$ has no further zeros. For the $ p$-wave (inter-sublattice) and $ f$-wave (intra-sublattice) pairing state, Eqs.~\eqref{inter} and \eqref{intra} provide a complete list of zeros within the BZ. 
Note that for general odd-parity pairing states, more zeros can appear whose position require a more complex analysis. Obviously, all such zeros require that both components 
of the pairing function $ \Phi_x (\mathbf{k}) $ and  $ \Phi_y (\mathbf{k}) $ vanish. 

\begin{figure}
	\includegraphics[width=0.90\columnwidth]{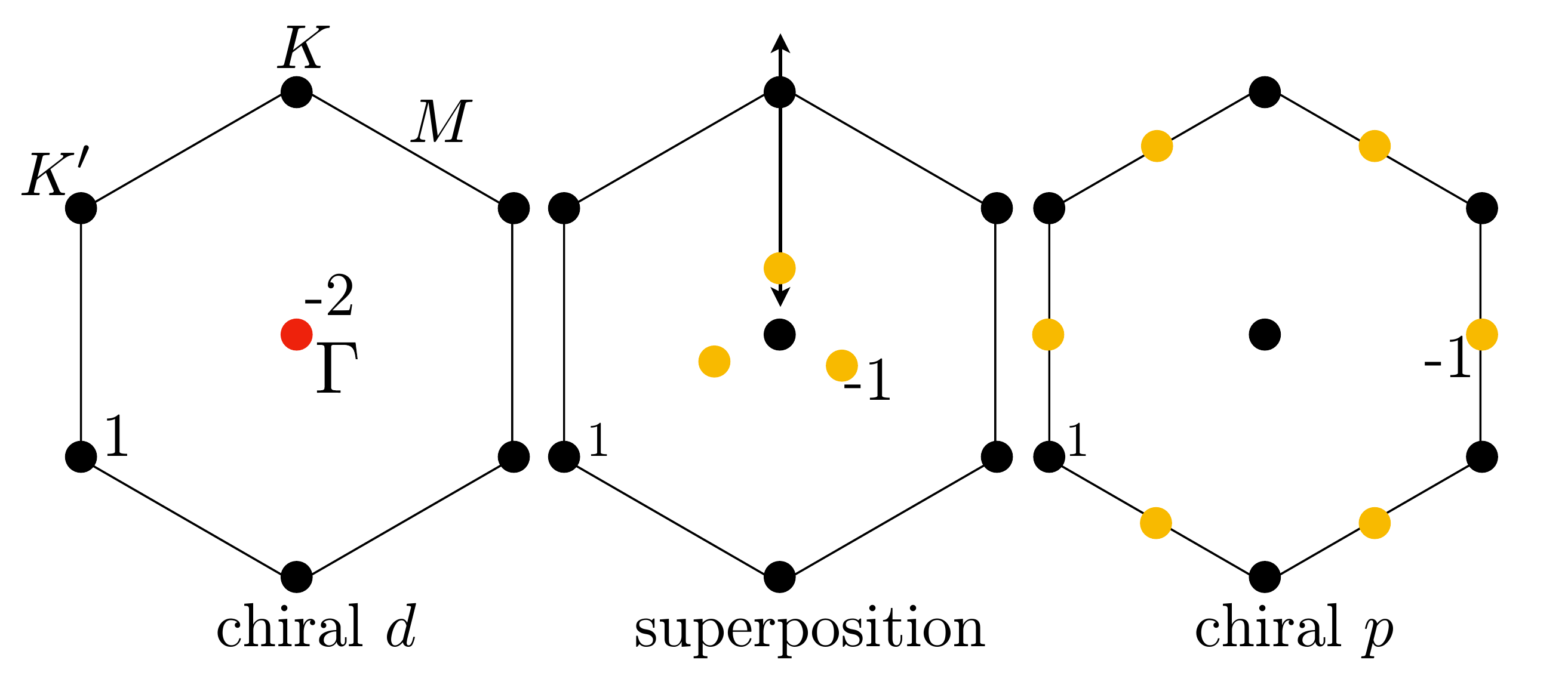}
	\caption{Zeros in the gap function and corresponding winding numbers for chiral $d$- and chiral $p$-wave gap functions, as well as a superposition of the two. Note that the situation sketched here corresponds to dominant $d$-wave pairing. The black arrow denotes the direction the zeros move when changing the ratio of the pairing channels.}
	\label{fig:trig}
\end{figure}

\section{Lattices with three-fold rotation axis} \label{sec:hexagonal}
In a hexagonal lattice with full $D_{6h}$ symmetry \cite{nandkishore:2012, kiesel:2012}, there are four two-dimensional representations, two even and two odd under inversion, allowing for both spin-singlet and spin-triplet chiral states. However, only one of each can be realized in the two-dimensional case. Using the primitive lattice vectors $\mathbf{T}_1 = (1, 0)$, and $\mathbf{T}_{2,3} = -(1/2, \mp \sqrt{3}/2)$, we can write them as
\begin{eqnarray}
	\Delta^d_{\textbf{k}} = \sum_n \omega^n \cos(\mathbf{T}_n\cdot \mathbf{k}) (i \sigma^y),\\
	\Delta^p_{\textbf{k}} = \sum_n \omega^n \sin(\mathbf{T}_n\cdot \mathbf{k})\sigma^z  (i \sigma^y),
\end{eqnarray}
with $\omega = \exp(2\pi i /3)$. These gap functions correspond, first, to a chiral $d$-wave and second, to a chiral $p$-wave order parameter. Due to the three-fold rotation, which maps $K$ ($K'$) to itself, but adds a non-trivial phase to the gap function, the order parameter has to vanish at the $K$ and $K'$ points for both order parameters. Additionally, the chiral $p$-wave order parameter has zeros on the $M$ points due to inversion symmetry, see Fig.~\ref{fig:trig}.

While these two order parameters can not mix in the hexagonal lattice and the zeros at the high-symmetry points are fixed, the situation is different when the lattice symmetry is reduced to $D_{3h}$ by making the two sublattices in the hexagonal lattice distinct, a situation  that is for example realized in the transition-metal dichalcogenides (TMDs)~\cite{ribak2019} or in a single layer of SrPtAs~\cite{nishikubo:2011, goryo:2012, fischer:2015a}. Lacking inversion symmetry, the system allows for combinations of the spin-singlet and spin-triplet order parameters. In particular, angular momentum is only defined modulo 3, such that the $l=1$ pairing state can mix with the one with $l=-2$. Starting from the chiral $d$-wave situation with $l=-2$, the zero at the $\Gamma$ point splits up into one zero with winding number $1$ and three with winding number $-1$. These latter zeros move towards the $K$ points, see Fig.~\ref{fig:trig}~\footnote{In this case, there are two non-degenerate bands with different relative phase in the order parameters, such that the other gap function has zeros moving towards the $K'$ points}. Depending on the Fermi surface topology, this can lead to several topological transitions until the zeros with $-1$ reach the $M$ points for a pure chiral $p$-wave state.

\section{Derivation of thermodynamic quantities} \label{sec: appendixB}
Here, we take a closer look at the derivation of 
the density of states, specific heat, superfluid density 
and thermal conductivity. We employ the Green's function 
formalism and linear response theory 
\cite{mineev1999introduction, langer1962js, ambegaokar1964theory, 
luttinger1964theory, schrieffer2018theory, nomura2005theory}.\par 
The density of states is in our framework \cite{mineev1999introduction} given by 
\begin{align}
	N(E) = -\frac{1}{\pi} \text{Im} \sum_{\mathbf{k}} G^{\text{ret}} (\mathbf{k}, E), \label{eqn: dos} 
\end{align} 
with $G^{\text{ret}}(\mathbf{k}, E)$ being the retarded Green's function, 
obtained by analytically continuing the frequencies 
to the real-axis, $i \omega_n \rightarrow E + i 0$. In a dirty system, $E$ 
is replaced by $\tilde{E} = E - \Sigma^{\text{ret}}(E)$. The explicit expression 
of the Green's functions is shown below.\par 
Moving on to the specific heat, we start with the generalization \cite{keller1988free} 
of the ground-state energy formula by Luttinger and Ward \cite{luttinger1960ground}
for an interacting electron system. The grand potential is given by 
\begin{align}
	\displaystyle \Omega_s = -T \sum_n \sum_{\mathbf{k}} &\Big\{ \log{(\tilde{\omega}_n^2 + \xi_{\mathbf{k}}^2 + |\Delta_{\mathbf{k}}|^2)}
	+ \Delta_{\mathbf{k}} F^{\dagger}(\mathbf{k}, i\omega_n)  \nonumber \\[1mm] 
	&+ \Sigma(i\omega_n) G(\mathbf{k}, i\omega_n) \Big\} + \Omega',
\end{align}
with $i\tilde{\omega}_n = i\omega_n - \Sigma(i\omega_n)$ and  
\begin{align}
	\displaystyle \Omega' = T \sum_{\nu} \sum_n \sum_{\mathbf{k}} \frac{1}{\nu} \Sigma_{\nu} (i \omega_n) G(\mathbf{k}, i\omega_n),
\end{align}
where $\Sigma(i\omega_n) = c T(i\omega_n) = \sum_{\nu} \Sigma_{\nu} (i \omega_n)$. 
To make sure that the sum over $n$ converges rapidly, we consider the difference 
between superconducting and normal state, $\Omega_s - \Omega_n$.\par 
Once the self-energy has been determined self-consistently, we calculate 
the specific heat difference through
\begin{align}
	\frac{C_s - C_n}{T} = -\frac{\partial^2 (\Omega_s - \Omega_n)}{\partial T^2} . \label{eqn: cv}
\end{align}
The specific heat in the normal state can now be added to the numerical solution 
of this equation to obtain $C_s/T$. Note that we have dropped the subscript in the main text.\par
Now let us take a look at the superfluid density $n_s$. 
In the clean system it is per definition~\cite{lee1997unusual}
equal to  
\begin{align}
	n_s(T) \equiv n_s(T=0) - n_n(T). 
\end{align}
Thus, it is enough to calculate $n_n$ by taking the real 
part of the current-current correlation function~\cite{schrieffer2018theory}, 
\begin{align}
	n_n(T) = - \lim_{\substack{\Omega\rightarrow 0}} \lim_{\substack{\mathbf{q}\rightarrow 0}} 
	\text{Re}\ \Pi_{\mu \nu}^{\text{ret}} (\mathbf{q}, \Omega) \delta_{\mu \nu}, 
\end{align}
whose definition is 
\begin{align}
	\Pi_{\mu \nu} (\mathbf{q}, \tau) = \braket{T_{\tau} j_{\mu}(\mathbf{q}, \tau) j_{\nu}(-\mathbf{q}, 0)}.
\end{align}
It incorporates the electrical-flux operator,
\begin{align}
	j_{\mu}(\mathbf{q}, \tau) = &\frac{1}{2} \underset{\tau ' \rightarrow \tau}{\lim} \sum_{\mathbf{k} \sigma} (v_{\mathbf{k} + \mathbf{q} \mu} + v_{\mathbf{k} \mu}) \nonumber \\[1mm]
		&\times c_{\mathbf{k} \sigma}^{\dagger} (\tau) c_{\mathbf{k} + \mathbf{q} \sigma} (\tau').
\end{align}
The $\mu$-component of the velocity, $v_{\mathbf{k} \mu}$, obeys the relation
\begin{align}
	v_{\mathbf{k} \mu} = \frac{\partial \xi_{\mathbf{k}}}{\partial k_{\mu}}. 
\end{align}
The final result then reads
\begin{align}
	n_n(T) = \frac{2}{V \beta} \sum_{\mathbf{k}} v_{\mathbf{k} \mu} v_{\mathbf{k} \nu} \sum_{n} \text{Re} &\Big[ (G(\mathbf{k}, i\omega_n))^2 \nonumber \\[1mm]
	+ F(\mathbf{k}, i\omega_n) F^{\dagger}(\mathbf{k}, i\omega_n) \Big], \label{eqn: ns}
\end{align}
where, according to Gor'kov's equations \cite{mineev1999introduction}, 
the Green's functions take the form 
\begin{align}
	G(\mathbf{k}, i\omega_n) &= -\frac{i\tilde{\omega}_n + \xi_{\mathbf{k}}}{\tilde{\omega}_n^2 + \xi_{\mathbf{k}}^2 + |\Delta_{\mathbf{k}}|^2}, 
	\label{eqn: eq4} \\[1mm]
	  F(\mathbf{k}, i\omega_n) &= \frac{\Delta_{\mathbf{k}}}{\tilde{\omega}_n^2 + \xi_{\mathbf{k}}^2 + |\Delta_{\mathbf{k}}|^2}, \label{eqn: eq7} \\[1mm] 
	  F^{\dagger}(\mathbf{k}, i\omega_n) &= \frac{\Delta_{\mathbf{k}}^{\ast}}{\tilde{\omega}_n^2 + \xi_{\mathbf{k}}^2 + |\Delta_{\mathbf{k}}|^2}. \label{eqn: eq8}
\end{align} 
Note that one has to use electron-hole symmetry, in other words a 
constant density of states, to obtain this 
specific structure of the Green's functions. 
Even if the system may not have this symmetry, we assume 
it here for the sake of simplicity. Note that the results 
are not affected by this assumption.\par  
Lastly, we consider the thermal conductivity $\kappa$. 
Again, linear response theory is applied. Our starting point is 
\begin{align}
	\kappa_{\mu \nu} = -\frac{1}{T} \underset{\Omega \rightarrow 0}{\lim}\ \frac{1}{\Omega}\ \underset{\mathbf{q} \rightarrow 0}{\lim}\ \text{Im}\ K^{\text{ret}}_{\mu \nu} (\mathbf{q}, \Omega), \label{eqn: eq5}
\end{align}
where $K^{\text{ret}}_{\mu \nu} (\mathbf{q}, \Omega)$ is the 
retarded thermal-flux correlation function. Without the analytical 
continuation of the bosonic Matsubara frequencies
,~$i\Omega_m\rightarrow\Omega + i0$, and Fourier transformation, 
it is given by
\begin{align}
	K_{\mu \nu} (\mathbf{q}, \tau) = \braket{T_{\tau} j^{\text{Th}}_{\mu}(\mathbf{q}, \tau) j^{\text{Th}}_{\nu}(-\mathbf{q}, 0)},  \label{eqn: eq6}
\end{align}
with $j_{\mu}^{\text{Th}}(\mathbf{q}, \tau)$ the thermal-flux operator, 
\begin{align}
	j^{\text{Th}}_{\mu}(\mathbf{q}, \tau) = &\frac{1}{2} \underset{\tau ' \rightarrow \tau}{\lim} \sum_{\mathbf{k}\sigma} \left( \frac{\partial}{\partial\tau} v_{\mathbf{k} + \mathbf{q} \mu} - \frac{\partial}{\partial\tau '} v_{\mathbf{k} \mu} \right) \nonumber \\[1mm]
	&\times c^{\dagger}_{\mathbf{k}\sigma}(\tau) c_{\mathbf{k} + \mathbf{q} \sigma}(\tau ') . 
\end{align}
After applying a standard mean-field approximation to Eq.~\eqref{eqn: eq6} and reformulating
everything in terms of Green's functions, we make use of the spectral representation of the 
latter, perform the Matsubara frequency summation and subsequently take the limits. 
In the end Eq.~\eqref{eqn: eq5} yields 
\begin{align}
	&\kappa_{\mu \nu} = \frac{1}{T^2} \int \frac{dE}{2\pi} \frac{E^2}{\cosh^2(E/2 T)}
	\frac{1}{V} \sum_{\mathbf{k}} v_{\mathbf{k}\mu} v_{\mathbf{k}\nu} \nonumber \\[1mm]
	&\times \Big[ (\text{Im}\ G^{\text{ret}}(\mathbf{k}, E))^2 - |\Delta_{\mathbf{k}}|^2(\text{Im}\ \bar{F}^{\text{ret}}(\mathbf{k}, E))^2 \Big]. \label{eqn: kappa}
\end{align}
The retarded Green's functions, which appear in this equation, are defined as
\begin{align}
	G^{\text{ret}}(\mathbf{k}, E) &= \frac{\tilde{E} + \xi_{\mathbf{k}}}{\tilde{E}^2 - \xi_{\mathbf{k}}^2 - |\Delta_{\mathbf{k}}|^2}, \label{eqn: eq9}\\[1mm] 
	  \bar{F}^{\text{ret}}(\mathbf{k}, E) &= -\frac{1}{\tilde{E}^2 - \xi_{\mathbf{k}}^2 - |\Delta_{\mathbf{k}}|^2}.
\end{align}

\bibliographystyle{apsrev4-1}
\bibliography{references}

\end{document}